\documentclass[10pt]{article}
\usepackage{amsfonts,amsmath,color,graphicx,stmaryrd,mathabx,amssymb,mathdots,esint,tikz,appendix,lscape}
\usepackage{bm} 
 
\DeclareMathAlphabet{\mathpzc}{OT1}{pzc}{m}{it}

\makeatletter
\@addtoreset{equation}{section}
\makeatother
\newtheorem{theorem}{Theorem}[section]
\newtheorem{conjecture}[theorem]{Conjecture}
\newtheorem{proposition}[theorem]{Proposition}
\newtheorem{lemma}[theorem]{Lemma}
\newtheorem{corollary}[theorem]{Corollary}

\newtheorem{example}{Example}[section]
\newtheorem{definition}[example]{Definition}

\newtheorem{remark}[example]{Remark}
\newtheorem{hypothesis}[example]{Hypothesis}
\def\br{\begin{remark}\rm\small}
\def\er{\end{remark}}
\def\bt{\begin{theorem}}
\def\et{\end{theorem}}
\def\bcj{\begin{conjecture}}
\def\ecj{\end{conjecture}}
\def\bd{\begin{definition}}
\def\ed{\end{definition}}
\def\bp{\begin{proposition}}
\def\ep{\end{proposition}}
\def\bl{\begin{lemma}}
\def\el{\end{lemma}}
\def\bc{\begin{corollary}}
\def\ec{\end{corollary}}
\def\bh{\begin{hypothesis}}
\def\eh{\end{hypothesis}}
\def\beaq{\begin{eqnarray}}
\def\eeaq{\end{eqnarray}}

\newcommand{\Tr}{\mathrm{Tr}\,}
\newcommand{\beq}{\begin{equation}}
\newcommand{\eeq}{\end{equation}}
\newcommand{\bea}{\begin{eqnarray}}
\newcommand{\eea}{\end{eqnarray}}

\newcommand{\dd}{\mathrm{d}}

\newcommand{\Res}{\mathop{\,\rm Res\,}}

\newcommand{\Ms}{\mathpzc{M}}
\newcommand{\Hs}{\mathpzc{H}}

\definecolor{rouge}{rgb}{0.84,0.18,0.07}
\definecolor{bleu}{rgb}{0.22,0.41,0.74}
\definecolor{vertf}{rgb}{0.08,0.46,0.07}
\usepackage[pdftex]{hyperref}
\hypersetup{colorlinks,urlcolor=vertf,citecolor=rouge,linkcolor=bleu,filecolor=black}

\begin{document}

\sloppy

\pagestyle{empty}
\addtolength{\baselineskip}{0.20\baselineskip}
\begin{center}

\vspace{26pt}

{\Large \textbf{Formal multidimensional integrals, stuffed maps, and topological recursion}}
\end{center}

%





\pagestyle{plain}
\setcounter{page}{1}


\vspace{26pt}
\begin{center}
\large{\textsl{Ga\"etan Borot}}\footnote{\href{mailto:gborot@mpim-bonn.mpg.de}{\textsf{gborot@mpim-bonn.mpg.de}}} \\
\vspace{0.2cm}
MPI f\"ur Mathematik \\
Vivatsgasse 7, 53111 Bonn, Germany \\
\end{center}

\vspace{26pt}

\begin{center}
\textbf{Abstract}
\end{center}

\noindent \textsf{We show that the large $N$ expansion in the multi-trace 1 formal hermitian matrix model is governed by the topological recursion of \cite{EOFg} with initial condition. In terms of a $1d$ gas of eigenvalues, this model includes -- on top of the squared Vandermonde -- multilinear interactions of any order between the eigenvalues. In this problem, the initial data $(\omega_1^0,\omega_2^0)$ of the topological recursion is characterized: for $\omega_1^0$, by a non-linear, non-local Riemann-Hilbert problem on the discontinuity locus $\Gamma$ to determine ; for $\omega_2^0$, by a related but linear, non-local Riemann-Hilbert problem on the discontinuity locus $\Gamma$. In combinatorics, this model enumerates discrete surfaces (maps) whose elementary 2-cells can have any topology -- $\omega_1^0$ being the generating series of disks and $\omega_2^0$ that of cylinders. In particular, by substitution one may consider maps whose elementary cells are themselves maps, for which we propose the name "stuffed maps". In a sense, our results complete the program of the "moment method" initiated in the 90s to compute the formal $1/N$ in the one hermitian matrix model.}

\section{Introduction}
\label{S1}

\subsection{Problem and main results}

It is well-known that the large $N$ expansion of the partition function and correlation functions in a $N \times N$ hermitian matrix model with measure:
\beq
\label{e0}\dd\mu(M) = \dd M\,e^{-N V(M)}
\eeq
is governed by a topological recursion \cite{ACM92,ACKM,ACKMe,E1MM}. This topological recursion takes a universal form and it goes far beyond the realm of matrix models. Eynard and Orantin have defined it axiomatically in the context of algebraic geometry \cite{EOFg}, and in this form, it enjoys many interesting properties (symplectic invariance, special geometry, WDVV equations, \ldots), and has appeared provably or experimentally in many problems of 2D enumerative geometry: the two hermitian matrix model \cite{EO2MM} and the chain of hermitian matrices \cite{CEO06}, topological string theory and Gromov-Witten invariants \cite{BKMP,BEMS,EMS,MuPen,Norbu,EOBKMP}, integrable systems \cite{BEdet,BETW,BEInt}, intersection numbers on the moduli space of curves \cite{EOwp,Ekappa,Einter}, asymptotic of knot invariants \cite{DiFuji2,BEknots,BEMknots}, \ldots

In this article, we extend the range of applicability of the topological recursion, by showing it governs (in the same universal form) the large $N$ expansion of formal hermitian matrix integrals based on the measure:
\beq
\label{e1} \dd\mu(M) = \dd M\,\exp\Big(\sum_{\substack{k \geq 1 \\ h \geq 0}} \frac{(N/t)^{2 - 2h - k}}{k!} \mathrm{Tr}\,T_k^h(M^{(k)}_{1},\ldots,M^{(k)}_k)\Big),
\eeq
where $M^{(k)}_i = \mathbf{1}_N \otimes \cdots \otimes M \otimes \cdots \mathbf{1}_N$ is a $k$-th tensor product where $M$ appears in $i$-th position, and $\mathbf{1}_N$ is the identity matrix. It induces the following measure of eigenvalues of $M$:
\beq
\dd\mu(\lambda_1,\ldots,\lambda_N) = \frac{\mathrm{Vol}(\mathrm{U}(N))}{N!(2\pi)^N}\prod_{i = 1}^N \dd\lambda_i \prod_{1 \leq i < j \leq N} (\lambda_i - \lambda_j)^2 \exp\Big(\sum_{\substack{k \geq 1 \\ h \geq 0}} \frac{(N/t)^{2 - 2h - k}}{k!} \sum_{i_1,\ldots,i_k = 1}^N T_k^{g}(\lambda_{i_1},\ldots,\lambda_{i_k})\Big).
\eeq
This is a generalization of the result obtained for arbitrary $2$-point interaction (i.e. $T_k^h \equiv 0$ whenever $(k,h) \neq (1,0),(2,0)$) in a recent work with Eynard and Orantin \cite{BEO}. As we explain in Section~\ref{S2}, the dependence in $N$ of the measure \eqref{e1} is the natural choice in order to have an expansion of topological nature.

We consider in the model \eqref{e1} the partition function:
\beq
Z = \mu[1] = \int \dd\mu(M),
\eeq
and the $n$-point correlation function:
\beq
\label{coer}W_n(x_1,\ldots,x_n) = \frac{\mu\Big[\prod_{j = 1}^n \mathrm{Tr}\,\frac{1}{x_j - M}\Big]_{c}}{\mu[1]},
\eeq
where the subscript $c$ stands for "cumulant" expectation value. In the context of formal matrix integrals, they have by construction a decomposition of the form:
\bea
Z & \propto &  \exp\Big(\sum_{g \geq 0} (N/t)^{2 - 2g}\,F^{g}\Big), \\
W_n(x_1,\ldots,x_n) & = & \sum_{g \geq 0} (N/t)^{2 - 2g - n}\,W_n^g(x_1,\ldots,x_n).
\eea
The precise definitions will be given in Section~\ref{S2}. Our main result are Theorems~\ref{thet} and \ref{thet2}, from which follows Theorem~\ref{tpoth}, which can be stated informally but with assumptions as follows:
\begin{proposition}
\label{P1} If the parameters of $T_k^h$ are tame (see Definition~\ref{tamde}), then all $W_n^g(x_1,\ldots,x_n)\dd x_1\cdots\dd x_n$ can be analytically continued to meromorphic $n$-forms on $\mathcal{C}^n$ for the same Riemann surface $\mathcal{C}$, can be computed by a recursion on $2g - 2 + n > 0$, which coincide up to an initial condition $\Phi_{n}^g(z,z_I)$ with the topological recursion of \cite{EOFg}. The initial data of this recursion is $W_1^0$ and $W_2^0$.
\end{proposition}
The tame condition is here the analog of an "off-criticality" condition in the context of random matrix theory.

\subsection{Motivations}

Beyond the effort to develop a complete theory of the topological recursion, let us motivate the study of models \eqref{e1}.

It is well-known that formal hermitian matrix integrals with measure \eqref{e0} enumerates maps, i.e. discrete surfaces obtained by polygonal faces with the topology of a disc along their edges. $V(x)$ is a generating series for the Boltzmann weight of such $2$-cells. The large $N$ expansion of the partition function and the correlation functions in these models collect maps of a given topology. Similarly, we show in Section~\ref{S2} that formal matrix integrals with measure \eqref{e1} enumerate \emph{stuffed maps}, i.e. maps obtained by gluing $2$-cells having the topology of a Riemann surface of genus $h$ with $k$ polygonal boundaries. $T_k^h(x_1,\ldots,x_k)$ is a generating series of such $2$-cells. Usual maps carrying self-avoiding loop configurations -- the so-called $O(n)$ model, introduced in a special case by \cite{K89} -- are equivalent to stuffed maps where the elementary cells may have the topology of a disc (usual faces) or of a cylinder (rings of faces carrying the loops). Usual maps with configuration of possibly intersecting loops can also be represented by stuffed maps. Therefore, the result of this article applies to many combinatorial models studied previously on usual random maps. In a sense, our results completes the program of the moment method \cite{ACM92,ACKM,ACKMe,EKOn} initiated in the context of $2$d quantum gravity to compute the large $N$ expansion in the one hermitian matrix model \eqref{e0}. Our result is that the same method, put in the form of the topological recursion \cite{EOFg}, applies to all multi-trace $1$ hermitian matrix models.

In \cite{GarMar}, Mari{\~{n}}o and Garoufalidis claim that, for any closed $3$ manifold obtained by filling in a knot $K$, the $\mathrm{U}(N)$ evaluation of the LMO invariants of a $3$-manifold can be computed from the $\mathrm{U}(N)$ Kontsevich integral of $K$, which is a formal $1$ hermitian matrix model, i.e. of the form \eqref{e1} for certain (in general non-explicit) weights $\mathbf{t}(\mathfrak{X})$. This indicates that the large $N$ expansion of those invariants should be described by a topological recursion. This will be the matter of a forthcoming work.

The convergent version of \eqref{e3} -- when it is well-defined -- describes a system of $N$ repulsive particles with position $\lambda_1,\ldots,\lambda_N$, which may have arbitrary $k$-point interactions. Such integrals frequently appear in the computation of correlation functions in quantum integrable systems, after applying Sklyanin's quantum separation of variables (see for instance \cite{KKN} in the example of the XXX spin chain and references therein).

\subsection{Outline}

We first define the formal model~\ref{e1} and the combinatorics of stuffed maps (Section~\ref{S2}), describe their nester structure in the case of planar maps and analyze some consequences (Section~\ref{S3}). Then, we write down the Schwinger-Dyson equations satisfied by the correlation functions \eqref{coer} (Section~\ref{S4}). They are equivalent to functional relations for generating series of stuffed maps, which can be given a bijective proof by Tutte's method. Their analysis (Section~\ref{S42}-\ref{S44}) shows that $W_n^g$ have the same type of monodromies around their discontinuity locus. More precisely, they satisfy a hierarchy of linear loop equations in the terminology of \cite{BEO} (Theorem~\ref{thet}). Then, the Schwinger-Dyson equations can be recast as quadratic loop equations (Section~\ref{S45}, Theorem~\ref{thet2}), and we can conclude in Section~\ref{S5} using the results of \cite{BEO} that $W_n^g$ for $2g - 2 + n > 0$ are given -- up to a shift for $(g,n) = (2,0)$ -- by the topological recursion (Theorem~\ref{tpoth}).

In practice, this reduces the problem of computing the sequence $(W_n^g)_{n,g}$ to the problem of computing $W_1^0$ and $W_2^0$. We show that $W_1^0$ is characterized by a scalar non-linear, non-local Riemann-Hilbert problem with a unknown jump locus $\Gamma$ (see \eqref{RHP1}), whereas $W_2^0$ is characterized by a related but linear, non-local Riemann-Hilbert problem on $\Gamma$ (see \eqref{RHP2}). In general, it seems hopeless to find the solution for $W_1^0(x)$ and $W_2^0(x_1,x_2)$ in closed form, but they can easily be obtained recursively as power series in the parameters of $T_k^h$.

The core of our computation is the analysis of the Schwinger-Dyson equation of Section~\ref{S5} to show linear and quadratic loop equations (Theorem~\ref{thet} and \ref{thet2}), and is relevant both for convergent and formal matrix integrals. It explains why the topological recursion holds in the same universal form in the class of models \eqref{e1}. The other technical details and assumptions are somewhat specific to the case of formal matrix integrals to which we restrict in this article. In the convergent matrix model, the assumptions and technical steps are of different nature and are more involved, because one needs first to justify the existence of a large $N$ expansion for an appropriate topology. In the more simple convergent model \eqref{e0}, the large $N$ asymptotic expansion were established in the one-cut case in \cite{APS01,BG11}, and in the multi-cut case in \cite{BGmulti} justifying the heuristics of \cite{BDE,Ecv} under natural assumptions on $V$. The generalization of this approach to the model \eqref{e1} seen as a convergent matrix model will be addressed in a subsequent work \cite{BGK}.

\section{The formal model}
\label{S2}
We recall the definition of \emph{formal} matrix integrals, and describe its underlying combinatorics in terms of stuffed maps. If $\mathbb{A}$ is a ring, and $\mathbf{t}$ is a collection of variables, $\mathbb{A}[[\mathbf{t}]]$ is the ring of formal series in $\mathbf{t}$ with coefficients in $\mathbb{A}$, whereas $\mathbb{A}[\mathbf{t}]$ is the polynomial ring of $\mathbb{A}$.

\subsection{Definition and notations}
\label{S21}
Let $\dd M$ be the Lebesgue measure on the space of $N \times N$ hermitian matrices $\mathcal{H}_N$, and $\mu_0$ be the Gaussian measure:
\beq
\dd\mu_0(M) = \dd M\,\exp\Big(-\frac{N\,\mathrm{Tr}\,M^2}{2t}\Big).
\eeq
Let $\mathbf{t} = (t_{\ell_1,\ldots,\ell_k}^h)_{\ell,k,h}$ a sequence of formal variables, assumed to be symmetric in $\ell_1,\ldots,\ell_k$. For any $k \geq 1$ and $h \geq 0$, we define a formal series depending on variables $\mathbf{p} = (p_{\ell})_{\ell \geq 1}$:
\beq
\widetilde{T}_k^h(\mathbf{p}) = \sum_{\ell_1,\ldots,\ell_k \geq 1} t_{\ell_1,\ldots,\ell_k}^h\,\prod_{i = 1}^{k} p_{\ell_i} \in \mathbb{C}[[\mathbf{p}]][[\mathbf{t}]].
\eeq
We introduce a exponential generating series:
\beq
\label{psi0}\psi(\mathbf{p}) = \exp\Big(\sum_{\substack{k \geq 1 \\ h \geq 0}} \frac{(N/t)^{2 - 2h - k}}{k!}\,\widetilde{T}_k^h(\mathbf{p})\Big) \in \mathbb{C}[[\mathbf{p}]][[\mathbf{t}]].
\eeq
Given a matrix $M$, we will specialize those variables to:
\beq
\label{psi1} p_{\ell}[M] = \frac{\mathrm{Tr}\,M^{\ell}}{\ell}.
\eeq
Then, we define the partition function $Z$ and the free energy $F$ as:
\bea
Z & = & \frac{\mu_0\big[\psi(\mathbf{p}[M])]}{\mu_0[1]} \in \mathbb{C}[[\mathbf{t}]],\nonumber \\
\label{Z} F & = & \ln Z \in \mathbb{C}[[\mathbf{t}]],
\eea
and the disconnected $n$-point correlation functions as:
\bea
\label{Wn} \overline{W}_n(x_1,\ldots,x_n) & = & \frac{1}{Z}\,\mu_0\Big[\psi(\mathbf{p}[M])\prod_{j = 1}^n \mathrm{Tr}\,\frac{1}{x_j - M}\Big] \in \mathbb{C}[[(x_j^{-1})_j]][[\mathbf{t}]].
\eea
If $I$ is a set with $n$ elements, we use the notation $W_n(x_I) = W_n((x_i)_{i \in I})$.
The connected $n$-point correlators $W_n(x_1,\ldots,x_n)$ can then be defined as the cumulant expectation values (instead of the moments) of $\mathrm{Tr}\,1/(x_j - M)$:
\beq
\label{Wnc}\overline{W}_n(x_1,\ldots,x_n) = \sum_{J \vdash \ldbrack 1,n \rdbrack} \prod_{i = 1}^{[J]} W_{|J_i|}(x_{J_i}),
\eeq
where the sum runs over partitions of $\ldbrack 1,n \rdbrack$, and $[J]$ denotes the number of subsets in the partition $J$.

\subsection{Multidimensional integrals}

Formally, if we disregard the dependence in $N$ that we chose in \eqref{psi0}, $\dd\mu_0(M)\psi(\mathbf{p}[M])$ is the most general measure on the space of $N \times N$ hermitian matrices which is invariant under conjugation. We may also diagonalize $M$ and consider the measure induced on its eigenvalues $\lambda_1,\ldots,\lambda_N$:
\beq
\label{e3}\boxed{(\dd\mu_0\cdot\psi)(\lambda_1,\ldots,\lambda_N)\,\,\propto\,\, \prod_{i = 1}^N\dd\lambda_i \prod_{1 \leq i < j \leq N} (\lambda_i - \lambda_j)^2 \exp\Big(\sum_{\substack{k \geq 1 \\ h \geq 0}} \frac{(N/t)^{2 - 2h - k}}{k!}\,\sum_{i_1,\ldots,i_k = 1}^N T_k^{h}(\lambda_{i_1},\ldots,\lambda_{i_k})\Big),}
\eeq
where we have introduced the formal series:
\beq
T_k^h(x_1,\ldots,x_k) = -\delta_{k,1}\delta_{h,0}\,\frac{x^2}{2t} + \sum_{m_1,\ldots,m_k \geq 1} \frac{t_{m_1,\ldots,m_k}^h}{m_1\cdots m_k}\,x_1^{m_1}\cdots x_k^{m_k}.
\eeq

\subsection{Stuffed maps}

\label{S22}

We now introduce the combinatorial model behind \eqref{e1}.

\begin{figure}[h!]
\begin{center}
\includegraphics[width=0.6\textwidth]{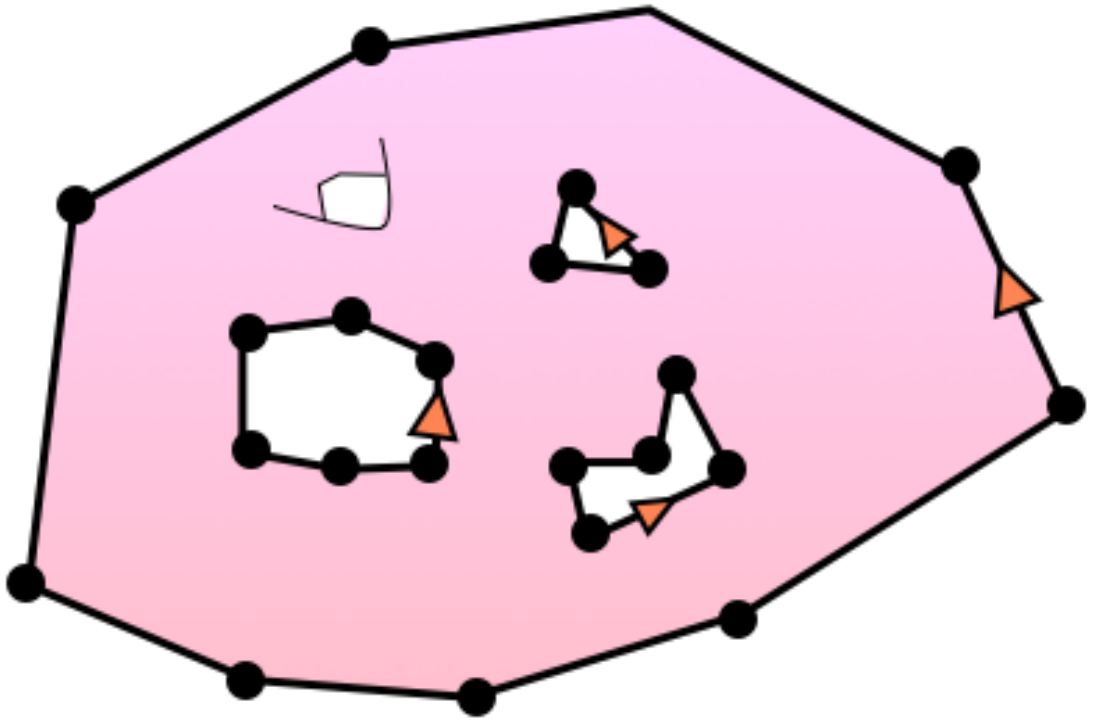}
\caption{\label{fcell} An elementary $2$-cell of topology $(k = 4,h = 1)$, with perimeters $\ell_1 = 9$, $\ell_2 = 6$, $\ell_3 = 5$ and $\ell_4 = 3$. The corresponding Boltzmann weight is $t^{1}_{9,6,5,3}$.}
\end{center}
\end{figure}

\begin{itemize}
\item[$\bullet$] An \emph{elementary 2-cell of topology $(k,h)$ and perimeters $(\ell_1,\ldots,\ell_k)$} is a topological, orientable, connected surface of genus $g$, with boundaries $B_i$ ($1 \leq i \leq k$) endowed with a set $V_i \subseteq B_i$ of $\ell_i \geq 1$ vertices (see Figure~\ref{fcell}). The connected components of $B_i\setminus V_i$ are considered as \emph{edges}.
\item[$\bullet$] A \emph{stuffed map of topology $(n,g)$ and perimeters $(\ell_1,\ldots,\ell_k)$} is a orientable, connected, discrete surface $\mathcal{M}$ of genus $g$, obtained from $n$ labeled rooted elementary $2$-cells with topology of a disc and perimeters $\ell_1,\ldots,\ell_n$, and from a finite collection of rooted unlabeled elementary $2$-cells, by gluing pairs of edges of opposite orientation. The labeled cells are considered as boundaries of the stuffed map, and the rooting on edges which do not belong to the boundary of $\mathcal{M}$ are forgotten after gluing. We denote $\mathbb{M}_{\ell_1,\ldots,\ell_n}^{g}$ this set of stuffed maps.
\item[$\bullet$] We say that a elementary $2$-cell (or a stuffed map) with boundaries is \emph{rooted} when a marked edge has been chosen on each boundary. By following the cyclic order, the rooting induces a labeling of the edges of the boundaries.
\end{itemize}
For instance, $(1,0)$ denotes the topology of a disc, $(2,0)$ denotes the topology of a cylinder, etc. A \emph{map} -- in the usual sense -- is a stuffed map made only of elementary $2$-cells with topology of a disc.

We assign a Boltzmann weight to stuffed maps in the following way:
\begin{itemize}
\item[$\bullet$] a weight $t$ per vertex.
\item[$\bullet$] a weight $t_{\ell_1,\ldots,\ell_k}^h$ per rooted elementary $2$-cell, depending on its topology $(k,h)$, and on its perimeters $\ell_1,\ldots,\ell_k$ in a symmetric way.
\item[$\bullet$] a symmetry factor $|\mathrm{Aut}\,\mathcal{M}|^{-1}$, where $\mathcal{M}$ is a stuffed map in which all constitutive elementary $2$-cells have been labeled and rooted, thus inducing a labeling for all edges. The identification of edges is thus represented by a permutation $\sigma$ which is a product of transposition of the edge labels. Aut $\mathcal{M}$ is the subgroup of permutations of elementary $2$-cells labels and rooting, for which we get the same stuffed map after identification of the edges according to $\sigma$ and forgetting all labels which do not decorate the boundary of $\mathcal{M}$.
\end{itemize}
By convention, the stuffed map consisting of only one vertex has $1$ boundary of length $0$, genus $0$, and thus receives a weight $1$. Out of a given finite collection of elementary $2$-cells, one can only construct a finite number of stuffed maps. This allows the definition:
\bea
F^g & = & \sum_{\mathcal{M} \in \mathbb{M}_{\emptyset}^{g}} \mathrm{weight}(\mathcal{M}) \in \mathbb{C}[[\mathbf{t}]], \\
W_n^g(x_1,\ldots,x_n) & = & \delta_{n,1}\delta_{g,0}\,\frac{t}{x_1} \\
& & + \sum_{\substack{\bm{\ell} = (\ell_1,\ldots,\ell_n) \\ \ell_1,\ldots,\ell_n \geq 1}} \Big[\prod_{j = 1}^n x_j^{-(\ell_j + 1)}\Big]\Big(\sum_{\mathcal{M} \in \mathbb{M}_{\bm{\ell}}^{g}} \mathrm{weight}(\mathcal{M})\Big) \in \mathbb{C}[[(x_j^{-1})_{j}]][[\mathbf{t}]],\nonumber 
\eea
where $t$ and $N$ are considered as variables, and $\mathbf{t} = (t_{\bm{\ell}}^h)_{\bm{\ell},h}$ as an infinite sequence of formal variables.

\subsection{Formal matrix model representation}
\label{S23}

Applying the standard techniques invented in \cite{BIPZ}, we quickly review the connection between the combinatorial model of \S~\ref{S22} and the formal matrix integrals of \S~\ref{S21}.

Given that $\mu_0$ is a Gaussian measure, Wick's theorem allows the computation of the coefficients of the formal series $\ln Z$, $\overline{W}_n$ and $W_n$ defined in \eqref{Z}-\eqref{Wn}-\eqref{Wnc} as sums over Feynman diagrams, which are fatgraphs. We claim that those fatgraphs are dual to stuffed maps. Indeed, we can represent a monomial $N^{2 - 2h - k}\Tr M^{\ell_1}\cdots \Tr M^{\ell_k}$ as a collection of $k$ fatvertices, with $\ell_i$ couples of ingoing edge/outgoing edge in cyclic order at the $i$-th fatvertex. The dual of this collection of fatvertices is a collection of $k$ polygonal faces, with perimeters $\ell_1,\ldots,\ell_k$, which form the boundaries of a single elementary $2$-cell of topology $(k,h)$. By construction, $\psi(\mathbf{p}[M]) \in \mathbb{C}[[\mathbf{t}]]$ defined in \eqref{psi0}-\eqref{psi1} is the generating series of collections of elementary $2$-cells, with a weight deduced from \S~\ref{S22}, and:
\begin{itemize}
\item[$\bullet$] an extra weight $(N/t)^{\chi}$ for each elementary $2$-cell with Euler characteristics $\chi$ ;
\item[$\bullet$] a symmetry factor corresponding $1/k!$ and $1/(\ell_1\cdots\ell_k)$ corresponding to labeling and rooting the boundaries of the elementary $2$-cells.
\end{itemize}
When we compute the $\mu_0$ expectation value of product of monomials, the Wick theorem mimics the gluing rules of elementary $2$-cells along edges of opposite orientations, and each pair of glued edges comes with a weight $t/N$. Each vertex in the stuffed map correspond in the dual picture of fatgraphs to a line on which flows a matrix index $i \in \ldbrack 1,N \rdbrack$, and thus receives an extra weight $N$. Taking into account the symmetry factors, the classical argument of t'Hooft \cite{tHooft} about Euler characteristics counting implies that the generating series of stuffed maps coincide with the correlation functions in the model \eqref{e1}:
\bea
\label{212} F & = & \sum_{g \geq 0} (N/t)^{2 - 2g}\,F^g, \\
\label{213} W_n(x_1,\ldots,x_n) & = & \sum_{g \geq 0} (N/t)^{2 - 2g - n}\,W_n^g(x_1,\ldots,x_n).
\eea
These equalities holds in $\mathbb{C}[[\mathbf{t}]]$ (resp. $\mathbb{C}[[(x_j^{-1})_j]][[\mathbf{t}]]$), meaning that for a given monomial in the formal variables $\mathbf{t}$, only finitely many $g$'s contribute to the sum.

If all Boltzmann weights $t,t_{\bm{\ell}}^h$ are non-negative, we may also define $F^g$ and the coefficients of $\prod_{j} x_j^{-(\ell_j + 1)}$ in $W_n^g$ as numbers in $[0,+\infty]$. If the latter happens to be finite for given non-negative values $t,t_{\bm{\ell}}$, they can also be defined as finite numbers for any real-valued weights $t'$ and $(t_{\bm{\ell}}^h)'$ so that $|t'| \leq t$ and $|(t_{\bm{\ell}}^h)'| \leq t_{\bm{\ell}}^h$.

\section{Disc generating series and substitution}
\label{S3}

\subsection{Substitution approach}
\label{S31}
\begin{figure}[h!]
\begin{center}
\includegraphics[width=0.8\textwidth]{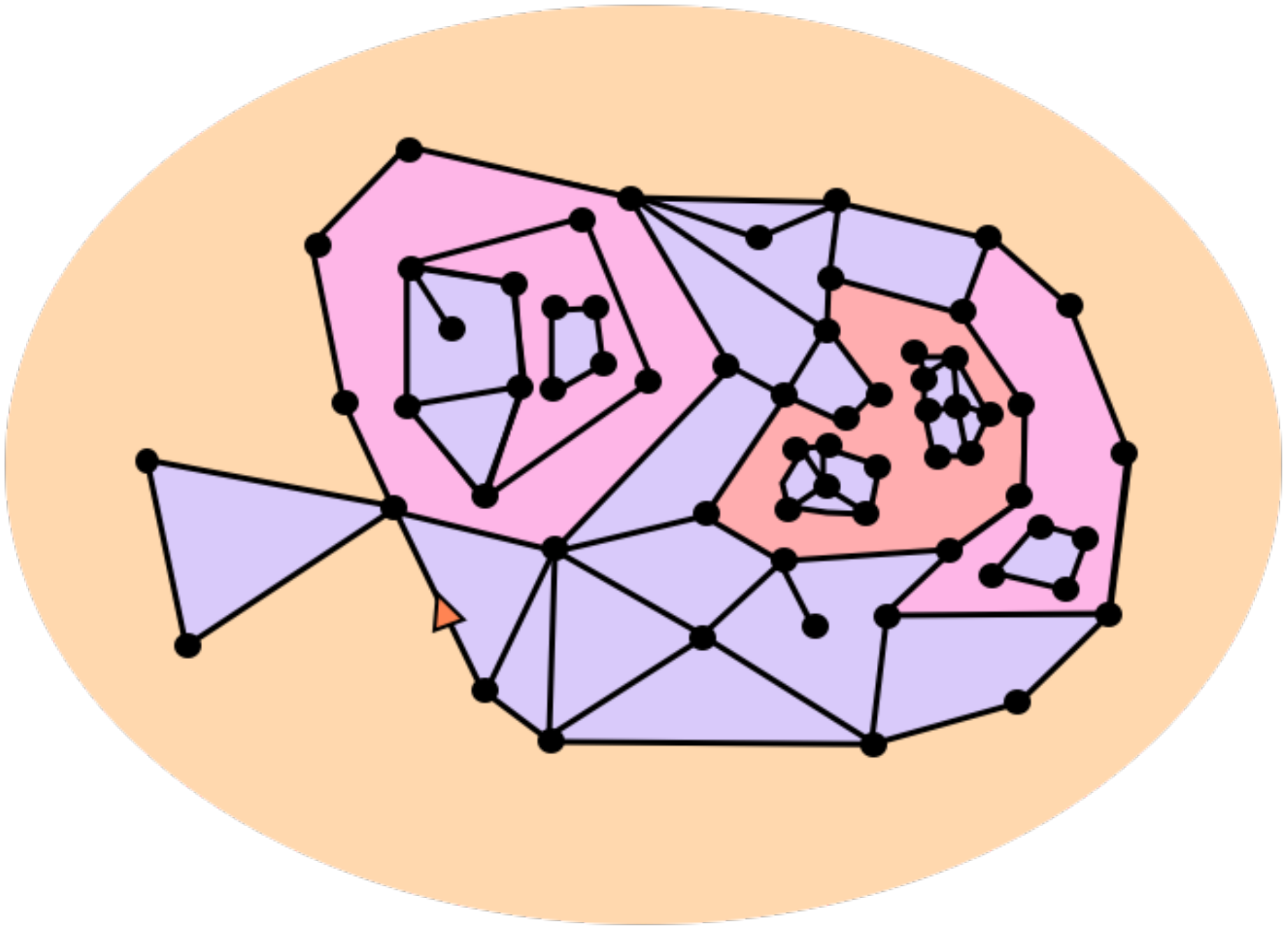}

\vspace{1cm}

\includegraphics[width=0.8\textwidth]{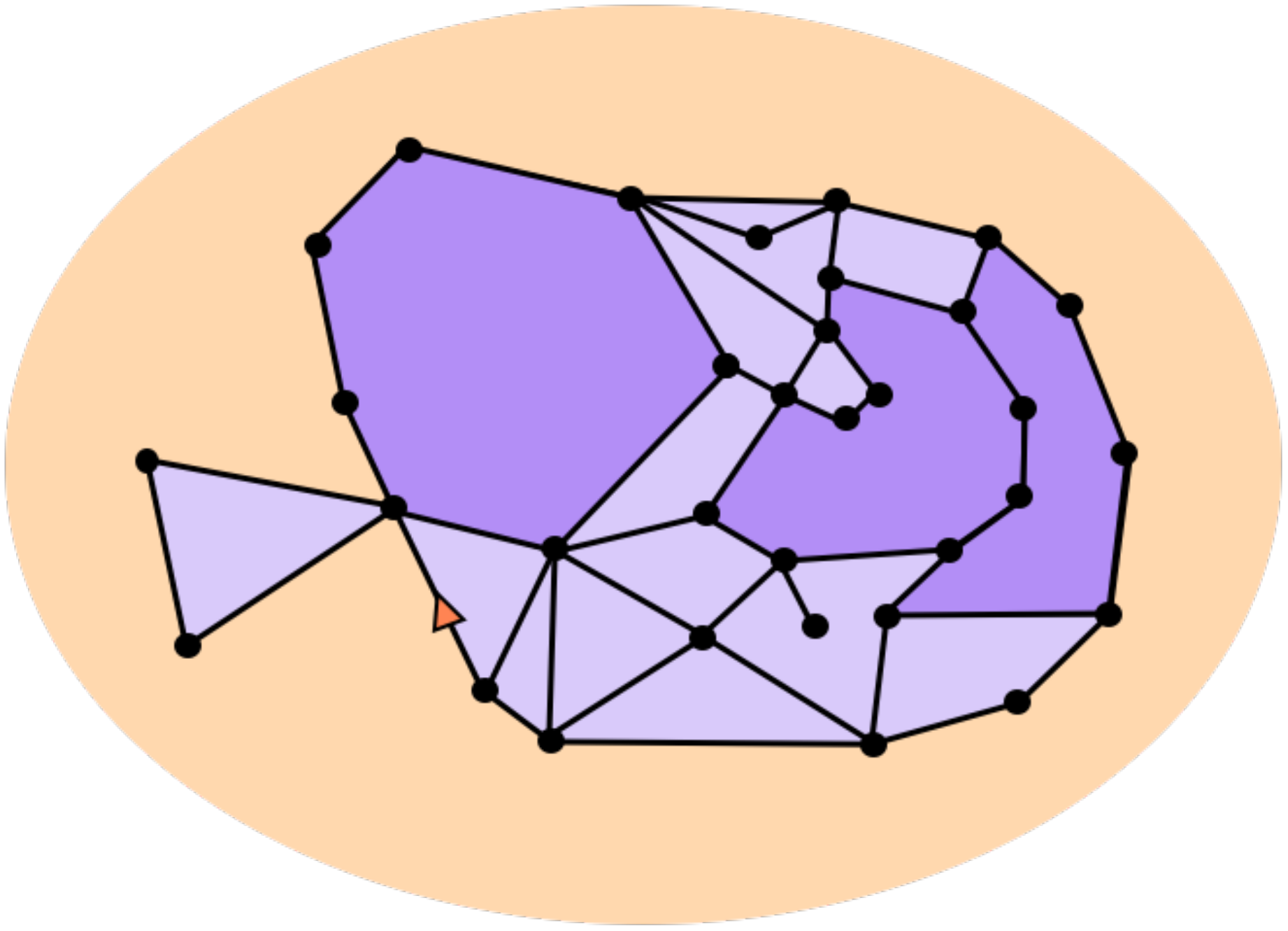}
\end{center}
\caption{\label{cf1} From top to bottom. First picture : a planar stuffed map with the topology of a disc. The orange arrow denote the root edge. We used different colors for elementary $2$-cells of different topology. The outer face -- peach color -- is the marked face. Second picture : the gasket of this stuffed map. The large faces appear in darker purple.}
\end{figure}

\begin{figure}[h!]
\begin{center}
\includegraphics[width=0.8\textwidth]{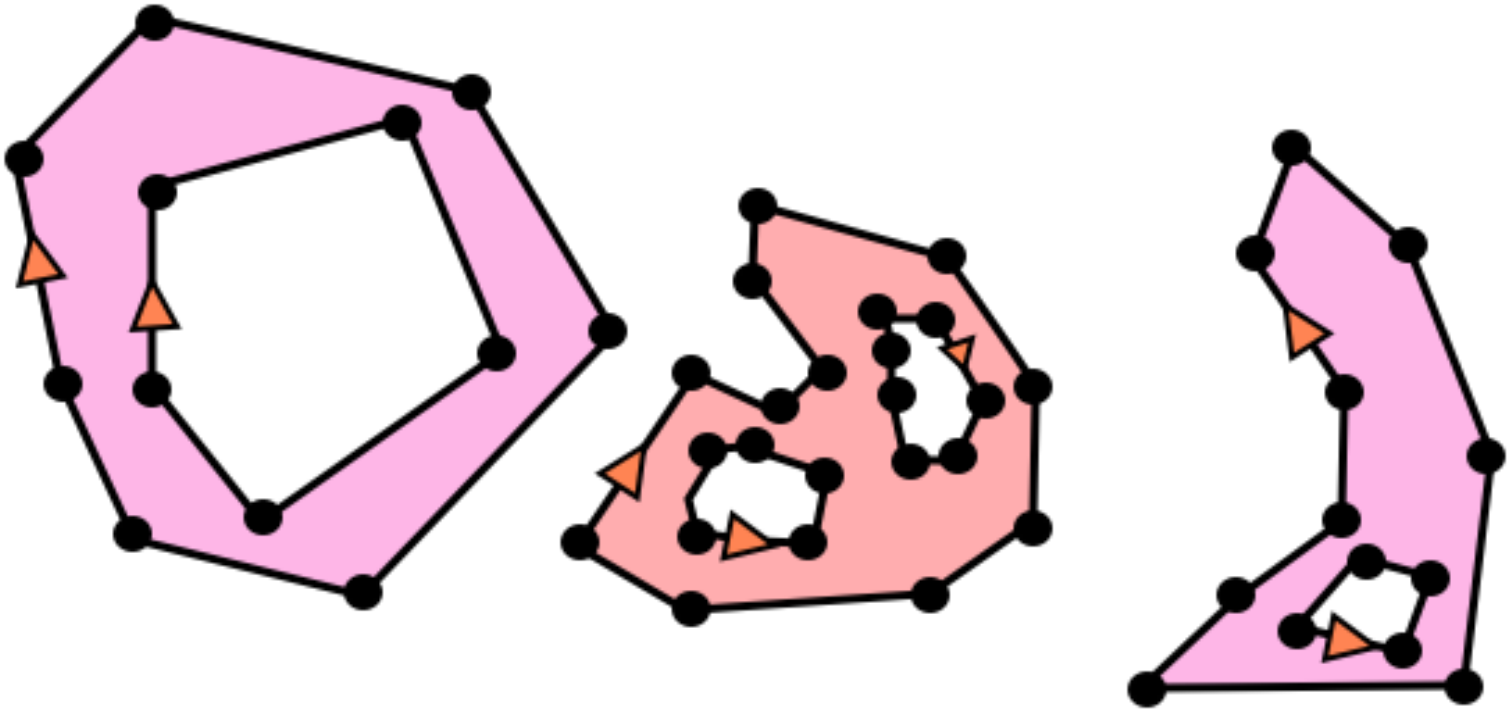}

\vspace{1cm}

\includegraphics[width=0.8\textwidth]{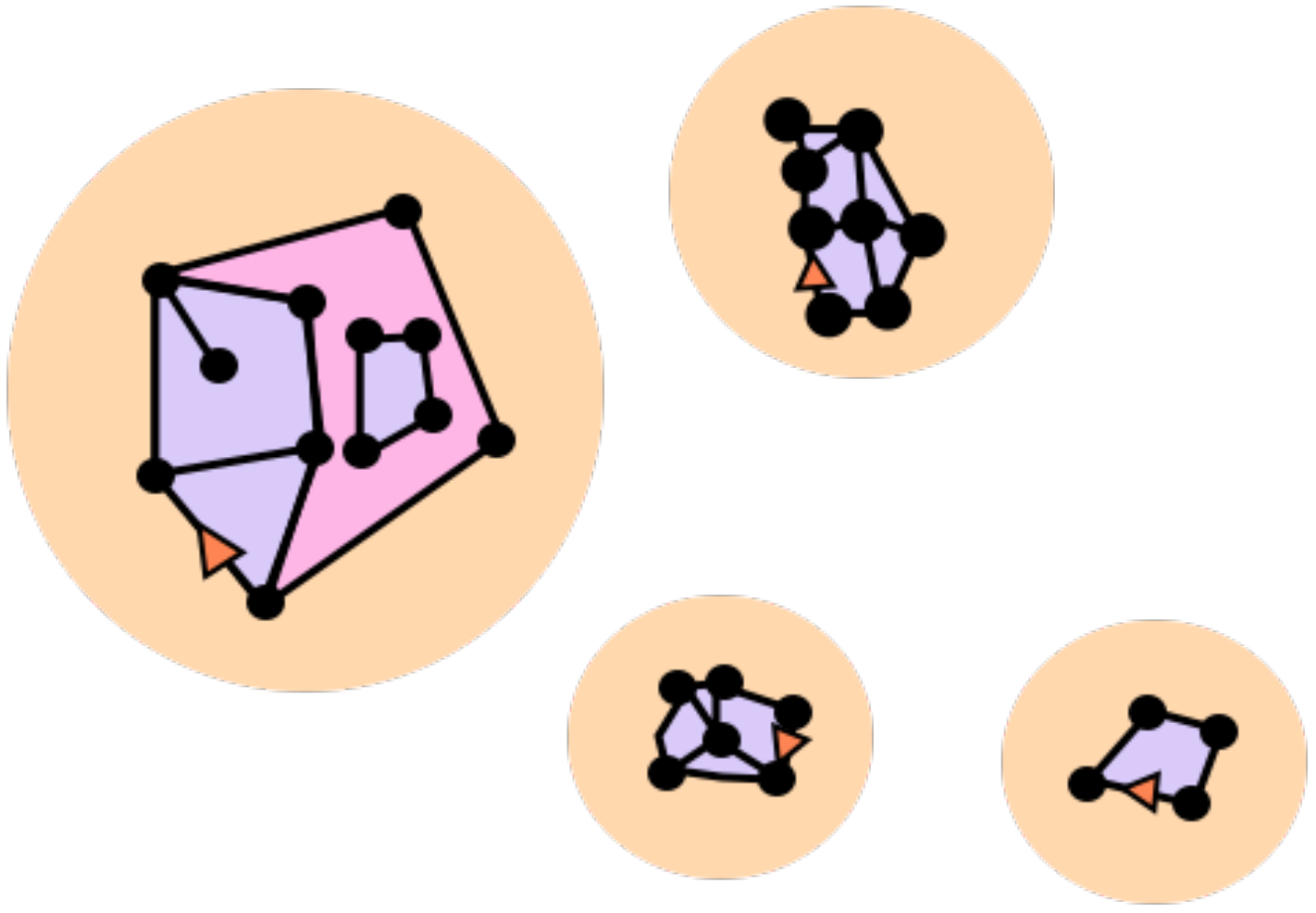}
\end{center}
\caption{\label{cf2} Third picture: the collection of its cement $2$-cells. Fourth picture: its chunks. The choice of root edges in both picture prescribes the way to glue them. To retrieve the map, we then have to forget the root edge on the boundaries of the chunks.}
\end{figure}

We first focus on planar stuffed maps $\mathcal{M}$ with topology of a disc, i.e. $(n,g) = (1,0)$. All their constitutive elementary $2$-cells must also be planar ($h = 0$), and if we remove one of them with $k$ boundaries, we end up with $k$ connected components. One of them contains the root edge on the boundary, and is called the \emph{exterior}, the other ones are tagged \emph{interior}. The existence of a notion of exterior and interior implies that planar stuffed maps have a nested structure, that we now describe (see Figure~\ref{cf1}-\ref{cf2}).

The \emph{gasket} $\mathcal{M}''$ of $\mathcal{M}$ is the map obtained by removing all elementary $2$-cells with $k \geq 2$ boundaries, keeping the connected component $\mathcal{M}'$ of the root edge in $\mathcal{M}$, and filling its holes having perimeter $m$ with new elementary $2$-cells with topology of a disc. We obtain in this way a usual map $\mathcal{M}''$ with topology of a disc, i.e. made only of elementary $2$-cells having the topology of a disc. Some of them were already $2$-cells in $\mathcal{M}$, and the other are called \emph{large faces}. The gasket $\mathcal{M}''$ does not contain all information about $\mathcal{M}$. It can be retrieved by specifying the configuration in the interior of $\mathcal{M}' \subseteq \mathcal{M}$. A hole in $\mathcal{M}'$ was created by the removal of a planar elementary $2$-cell with $k \geq 2$ boundary, which we call \emph{cement} $2$-cell. Since $\mathcal{M}$ is planar, distinct holes were created by the removal of distinct $2$-cells. The interior of a cement $2$-cell can be
seen as stuffed maps with topology of a disc, which we call \emph{chunks}.

We choose an arbitrary procedure to root the large faces of the gasket: among the points of a large face $\gamma$ which are the closest (for graph distance in $\mathcal{M}''$) to the point at the origin of the boundary of $\mathcal{M}''$, we choose the one $o'$ reached by the leftmost geodesic, and we root $\gamma$ on the edge with origin $o'$. We also root the corresponding edge on the cement $2$-cell filling this large face. 

Conversely, given a gasket, cement $2$-cells rooted on all their boundaries and rooted chunks, we can reconstruct the map $\mathcal{M}$ by gluing. The root edge on the chunks and the root edge on the corresponding boundary of a cement $2$-cell are identified in this process. This gluing is surjective, and if $m_i$ denote the sequence of perimeters of the chunks, it is actually $\prod_{i} m_i$ to $1$, since we must forget the roots on the boundaries of the chunks.

\subsection{Functional relation between generating series}
\label{S32}
Let $G_\ell[t,\mathbf{t}^0] \in \mathbb{C}[[\mathbf{t}^0]]$ be the generating series of stuffed maps with topology of a disc and perimeter $\ell$, and $G_{\ell}^{\mathrm{usual}}[t,\mathbf{t}^0] \in \mathbb{C}[[\mathbf{t}^0]]$ the analog for usual maps, obtained from $G_{\ell}[t,\mathbf{t}^0]$ by setting all $t_{m_1,\ldots,m_k}^0$ with $k \geq 2$ to zero. The bijection we described implies the simple functional relation:
\beq
\label{mm}\boxed{G_{\ell}[t,\mathbf{t}^0] = G_{\ell}^{\mathrm{usual}}[t,\bm{\tau}(t,\mathbf{t}^0)].}
\eeq
The right-hand side is the generating series for the gasket, which is a usual map whose $2$-cells were either present in the initial map (weights $\mathbf{t}^0$), or are large faces in which we glue a cement planar $2$-cell with $k \geq 2$ boundaries, and $(k - 1)$ stuffed maps with topology of a disc. We are cautious to add a symmetry factor to forget the roots on the chunks:
\bea
\tau_m(t,\mathbf{t}^0) & = & t_m^0 + \sum_{k \geq 2} \frac{1}{(k - 1)!} \sum_{m_2,\ldots,m_k \geq 1} \frac{t_{m,m_2,\ldots,m_k}^0}{m_2\cdots m_k} \prod_{i = 2}^k G_{m_i}[t,\mathbf{t}^0] \nonumber \\
& = & \sum_{k \geq 1} \frac{1}{(k - 1)!} \sum_{m_2,\ldots,m_k \geq 1} \frac{t_{m,m_2,\ldots,m_k}^0}{m_2\cdots m_k} \prod_{i = 2}^k G_{m_i}[t,\mathbf{t}^0]
\eea
$\bm{\tau}(t,\mathbf{t}^0)$ represents a sequence of effective face weights allowing to enumerate planar stuffed maps as planar usual maps. The properties of the generating series of planar usual maps $G_{\ell}^{\mathrm{usual}}[t,\bm{\tau}]$ are well known, and by \eqref{mm} they can be transferred to the generating series of stuffed maps $G_{\ell}[t,\mathbf{t}^0]$. 

We recall the definition of admissible weights \cite{BBG2}. For usual maps, a vertex weight $t$ and a sequence of non-negative face weights $\bm{\tau} = (\tau^0_1,\tau^0_2,\tau_3^0,\ldots)$ is admissible is for any $\ell \geq 1$, the generating series of pointed rooted maps with topology of a disc $t\partial_t G_{\ell}^{\mathrm{usual}}[t,\bm{\tau}]$ is finite. We also say that real-valued $t,\bm{\tau}$ are admissible if $|\bm{\tau}| = (|\tau^0_1|,|\tau^0_2|,|\tau^0_3|,\ldots)$ is admissible. For stuffed maps, we will say that a vertex weight $t$ and a sequence of elementary $2$-cells weights $\mathbf{t}^0$ is admissible if the effective face weights $\bm{\tau}(t,\mathbf{t}^0)$ are admissible. The admissibility condition is not empty:

\begin{lemma}
\label{L3}
When only a finite number of $t_{m_1,\ldots,m_k}^0$ are non-zero and have given values, there exists $t_c > 0$ so that, for any $|t| < t_c$, the weights $t,\mathbf{t}^0$ are admissible.\end{lemma}
\textbf{Proof.} It correspond to have usual maps with bounded face degree. The existence of $t_c > 0$ in this case is well-known, and can easily be deduced from \cite[Section 6]{BBG2}. \hfill $\Box$ 

\vspace{0.2cm}

As a consequence of \cite{BBG2}, for stuffed maps, we obtain a planar $1$-cut lemma and a functional relation:
\begin{lemma}
\label{L1}If $t,\mathbf{t}^0$ is a sequence of admissible weights for planar elementary $2$-cells, then \mbox{$G_{\ell}[t,\mathbf{t}^0] < \infty$} for all $\ell \geq 1$. The formal Laurent series:
\beq
W_1^0(x) = \frac{t}{x} + \sum_{\ell \geq 1} \frac{G_{\ell}[t,\mathbf{t}^0]}{x^{\ell + 1}}
\eeq
is the Laurent expansion at $\infty$ of a holomorphic function in $\mathbb{C}\setminus \Gamma_{t,\mathbf{t}^0}$, where $\Gamma_{t,\mathbf{t}^0}$ is a segment of the real line. Besides, $W_1^0(x)$ has limits from above and from below on $\Gamma_{t,\mathbf{t}^0}$, remains bounded, and
\beq
\rho(x) = \frac{W_1^0(x - {\rm i}0) - W_1^0(x + {\rm i}0)}{2{\rm i}\pi}
\eeq
assumes positive values at interior points of $\Gamma_{t,\mathbf{t}^0}$, and vanishes at the edges \hfill $\Box$
\end{lemma}
Let us introduce the generating series $\widetilde{V}_1^0(x)$ of planar elementary $2$-cells, whose boundaries are all glued to stuffed maps with topology of a disc, except one boundary which receives a weight $x^{\ell}$ when it has perimeter $\ell$. We also include a shift and a sign for convenience:
\bea
\widetilde{V}_1^0(x) & = & -\frac{x^2}{2t} + \sum_{\ell \geq 1} \sum_{m_1,m_2,\ldots,m_r \geq 1} \frac{t_{m_1,m_2,\ldots,m_k}}{m_1\cdots m_k}\,x^{m_1}\prod_{j = 2}^k G_{m_j}[t,\mathbf{t}^0] \nonumber \\
& = & \sum_{k \geq 1} \oint T_k^0(x,\xi_2,\ldots,\xi_k)\,\prod_{j = 2}^k \frac{\dd\xi_j\,W_1^0(\xi_j)}{2{\rm i}\pi}.
\eea

\begin{lemma}
\label{L2} If $t,\mathbf{t}^0$ is a sequence of admissible weights for planar elementary $2$-cells, there exists an open disc $\mathcal{D}_{t,\mathbf{t}^0}$ centered on $0$ and containing the interior of $\Gamma_{t,\mathbf{t}^0}$ so that the formal series $T_k^0(x_1,\ldots,x_k)$ defines a holomorphic function for $(x_1,\ldots,x_k) \in \mathcal{D}^{k}_{t,\mathbf{t}^0}$, and $\widetilde{V}_1^0(x)$ defines a holomorphic function for $x \in \mathcal{D}_{t,\mathbf{t}^0}$. Besides, for any $x$ in the interior of $\Gamma_{t,\mathbf{t}^0}$,
\beq
\label{RHP1}W_1^0(x + {\rm i}0) + W_1^0(x - {\rm i}0) + \partial_x \widetilde{V}_1^0(x) = 0.
\eeq
\hfill $\Box$
\end{lemma}
\eqref{RHP1} is a non-linear and non-local Riemann-Hilbert problem for $W_1^0$, with unknown discontinuity locus $\Gamma$. We will discuss in \S~\ref{S36} the unicity of its solution.

\subsection{Holomorphic functions with a cut}

\begin{definition}
If $U$ is an open set of the Riemann sphere, we define $\Ms^0(U)$ (resp. $\Hs^0(U)$) the space of meromorphic (resp. holomorphic) functions on $\Omega$. An open set $U \subseteq \widehat{\mathbb{C}}\setminus\Gamma$ which is a neighborhood of $\Gamma$ is called an \emph{exterior neighborhood} of $\Gamma$.
\end{definition}
Let us introduce a generating series of planar elementary $2$-cells, in which all but two boundaries are glued to stuffed maps with topology of a disc:
\bea
\label{tildeR}\widetilde{R}(x,y) & = & \sum_{k \geq 2} \frac{1}{(k - 1)!}  \oint  T_k^0(x,y,\xi_3,\ldots,\xi_k)\,\prod_{j = 3}^k \frac{\dd\xi_j\,W_1^0(\xi_j)}{2{\rm i}\pi}, \\
\label{Rdef} R(x,y) & = & \sum_{k \geq 2} \frac{1}{(k - 2)!} \oint T_k^0(x,y,\xi_3,\ldots,\xi_k)\,\prod_{j = 3}^k \frac{\dd\xi_j\,W_1^0(\xi_j)}{2{\rm i}\pi}.
\eea
The symmetry factor is the only difference between the two expressions. In order to work with analytic functions rather than formal series, we need slightly stronger assumptions.
\begin{definition}
\label{comp}
\begin{itemize}
\item[$\phantom{bla}$]
\item[$\bullet$] We say that admissible weights $t,\mathbf{t}^0$ are \emph{off-critical} when $\partial_{x} T_1^0(x)$ is holomorphic in an open neighborhood of $\Gamma_{t,\mathbf{t}^0}$. 
\item[$\bullet$]  We say that a sequence $(\tau_{\mathbf{m}})_{\mathbf{m}}$ is \emph{regular} when the formal series
$$
\sum_{m_1,\ldots,m_r \geq 1} \frac{\tau_{m_1,\ldots,m_k}}{m_1\cdots m_k}\,x_1^{m_1}\cdots x_r^{m_k} \in \mathbb{C}[[x_1,\ldots,x_r]]
$$
defines a holomorphic function in $\mathcal{D}^{r}$, where $\mathcal{D}$ is an open neighborhood of $\Gamma_{t,\mathbf{t}^0}$.
\item[$\bullet$] We say that admissible weights $t,\mathbf{t}^0$ are \emph{completely regular} when it is admissible, off-critical, $(\mathbf{t}^0_{m_1,\ldots,m_k})_{\mathbf{m}}$ is regular for any $k \geq 1$, and moreover $R(x,y)$ is holomorphic in $\mathcal{D}^2$, where $\mathcal{D}$ is an open neighborhood of $\Gamma_{t,\mathbf{t}^0}$.
\end{itemize}
\end{definition}
Let $t,\mathbf{t}^0$ be completely regular weights, $U$ be an open exterior neighborhood of $\Gamma$, and $U'$ be an open neighborhood of $\Gamma$. We can define a linear operator $\widetilde{\mathcal{O}}\,:\,\Hs^0(U) \rightarrow \Hs^0(U')$ by
\beq
\label{Otdef}\widetilde{\mathcal{O}}\phi(x) = \oint_{\Gamma} \partial_{x}\widetilde{R}(x,\xi)\,\phi(\xi),\qquad \mathcal{O}\phi(x) = \oint_{\Gamma} \partial_{x} \widetilde{R}(x,\xi)\,\phi(\xi).
\eeq
Besides, we also define the expressions:
\beq
\label{Sde}\mathcal{S}\phi(x) = \phi(x + {\rm i}0) + \phi(x - {\rm i}0),\qquad \Delta \phi(x) = \phi(x + {\rm i}0) - \phi(x - {\rm i}0).
\eeq
Eqn.~\ref{RHP1} can be rewritten: for any interior point $x$ of $\Gamma_{t,\mathbf{t}^0}$,
\beq
\label{acco}\mathcal{S}W_1^0(x) + \widetilde{\mathcal{O}}W_1^0(x) + \partial_x T_1^0(x)= 0.
\eeq
Since the two last terms are holomorphic in a neighborhood of $\Gamma$ and $W_1^0(x)$ remains bounded, we deduce:
\begin{lemma}
\label{L4} If $t,\mathbf{t}^0$ are completely regular, $W_1^0(x)$ can be decomposed, at $\alpha = a,b$ the edges of $\Gamma_{t,\mathbf{t}^0}$, as $h_1(x) + h_2(x)\sqrt{x - \alpha}$ where $h_1,h_2$ are holomorphic in a neighborhood of $\alpha$. \hfill $\Box$
\end{lemma}

\subsection{Analytic continuation}
\label{S34}
\begin{figure}[h!]
\begin{center}
\includegraphics[width=0.99\textwidth]{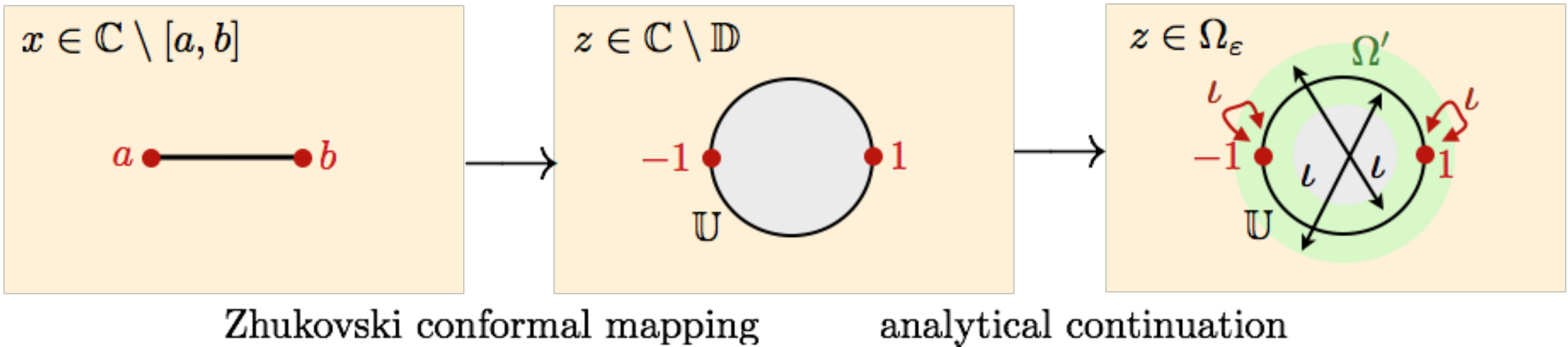}
\end{center}
\caption{\label{anana} Analytic continuation in the $z$-plane of functions of $x$ via \eqref{joukova}, $\iota(z) = 1/z$.}
\end{figure}

We start with some preliminaries about analytical continuation. Let $\Gamma = [a,b]$ be a segment of $\mathbb{R}$. The domain $\widehat{\mathbb{C}}\setminus \Gamma$ can be mapped conformally to the exterior of the unit disc $\overline{\mathbb{D}}$ by the Zhukovski map (see Figure~\ref{anana}):
\beq
\label{joukova}x(z) = \frac{a + b}{2} + \frac{a - b}{4}\Big(z + \frac{1}{z}\Big) \qquad \Longleftrightarrow \qquad x(z) = \frac{2}{a - b}\Big(x - \frac{a + b}{2} + \sqrt{(x - a)(x - b)}\Big).
\eeq
The image of the unit circle $\mathbb{U}$ by $x$ is $[a,b]$. We have a holomorphic involution $\iota(z) = 1/z$, which has $z(a) = 1$ and $z(b) = -1$ as fixed points. We have a notion of exterior or interior neighborhoods of $\mathbb{U}$. From now on, we prefer to work with differential forms rather than functions.

\begin{definition}
If $\Omega \subseteq \widehat{\mathbb{C}}$ is an open set, $\Ms(\Omega)$ (resp. $\Hs(\Omega)$) is the space of meromorphic (resp. holomorphic) $1$-forms in $\Omega$.
\end{definition}
If $\phi$ is a holomorphic function in an exterior neighborhood $U$ of $\Gamma$, upon multiplication by $\dd x$ it defines an element $\varphi \in \Hs(\Omega)$, where $\Omega$ is the exterior neighborhood of $\mathbb{U}$ such that $x(\Omega) = U$. Similarly, if $\phi$ is a holomorphic function in a neighborhood $U'$ of $\Gamma$, it defines an element $\varphi \in \Hs(\Omega')$ with $\Omega' = z(U')$ is an open neighborhood of $\mathbb{U}$ stable under $\iota$, and such that $\varphi(z) = \varphi(\iota(z))$. We can thus define linear operators $\mathcal{O},\widetilde{\mathcal{O}}\,:\,\Hs(\Omega) \rightarrow \Hs(\Omega')$ upgrading \eqref{Otdef} to $1$-forms in the $z$-plane. Besides, if $\Omega'$ is an open neighborhood of $\mathbb{U}$ stable under $\iota$, we may define $\mathcal{S},\Delta\,:\,\Ms(\Omega') \rightarrow \Ms(\Omega')$ by:
\beq
\label{Sde2}\mathcal{S}\varphi(z) = \varphi(z) + \varphi(\iota(z)),\qquad \Delta\varphi(z) = \varphi(z) - \varphi(\iota(z)).
\eeq
The restriction of \eqref{Sde2} to $z \in \mathbb{U}$, pulled-back by the map $z$, agrees with the definition \eqref{Sde} in terms of boundary values on $\Gamma$. We will apply repeatedly the following principle:
\begin{lemma}
\label{L5} Let $U$ be an exterior neighborhood of $\Gamma$, and $\phi \in \Hs^0(U)$. Assume that $\phi$ has boundary values on the interior of $\Gamma = [a,b]$, that for any $\alpha \in \{a,b\}$ there exists an integer $r$ so that $\phi(x)(x- \alpha)^{r_{\alpha}/2}$ remains bounded when $x \rightarrow \alpha$, and that $\mathcal{S}\phi(x)$ can be analytically continued as a holomorphic function in a neighborhood of $\Gamma$. Then, $\varphi(z) = \phi(x(z))\dd x(z)$, initially a holomorphic $1$-form in the exterior neighborhood $\Omega$ of $\mathbb{U}$ such that $x(\Omega) = U$, can be analytically continued to a meromorphic $1$-form in an open neighborhood $\Omega'$ of $\mathbb{U}$ which is stable under $\iota$. For $\alpha = \pm 1$, if $r_{\alpha} \geq 2$, it has a pole of order atmost $r_{\alpha} - 1$ at $z = \alpha$. \hfill $\Box$
\end{lemma}

We assume that $t,\mathbf{t}^0$ are completely regular, and that the generating series of stuffed maps with topology of a disc $W_1^0(x)$ is known. It is considered as a holomorphic function on $\mathbb{C}\setminus\Gamma$ for some segment $\Gamma \subseteq \mathbb{R}$. According to \eqref{acco}, $\mathcal{S}W_1^0(x)$ can be analytically continued as a holomorphic function in a neighborhood of $\Gamma$, and thanks to Lemma~\ref{L4}, we can apply Lemma~\ref{L5} to define:
\beq
\mathcal{W}_1^0(z) = W_1^0(x(z))\dd x(z)
\eeq
as a meromorphic $1$-form in:
\beq
\label{Ome}\Omega_{\varepsilon} = \big\{z \in \widehat{\mathbb{C}},\quad |z| > 1- \varepsilon\big\}
\eeq
 for some $\varepsilon > 0$. Its only singularity is a simple pole with residue $-t$ at $z = \infty$, and it satisfies for any $z \in \Omega_{\varepsilon}\cap\iota(\Omega_{\varepsilon})$:
\beq
\boxed{\mathcal{S}\mathcal{W}_1^0(z) + \widetilde{\mathcal{O}}\mathcal{W}_1^0(z) + \dd_{z} T_1^0(x(z)) = 0.}
\eeq

\subsection{The master operator}
\label{S35}

The operator $\mathcal{O}$ will play an important role in the study of higher topologies, let us recall its definition in the realm of $1$-forms:
\beq
\label{Oop}\mathcal{O}\varphi(z) = \sum_{k \geq 2} \frac{1}{(k - 2)!} \oint_{\mathbb{U}^{k - 1}} \dd_{z} T_k^0(x(z),x(\zeta_2),\ldots,x(\zeta_k))\,\varphi(\zeta_2) \prod_{j = 3}^k \mathcal{W}_1^0(\zeta_j).
\eeq
If $\Omega'$ is a neighborhood of $\mathbb{U}$ stable under $\iota$, we want to study the space of solutions $\varphi \in \Ms(\Omega')$ of:
\beq
\label{solsa}\mathcal{S}\varphi(z) + \mathcal{O}\varphi(z) = 0.
\eeq
We start with a result of unicity. The unicity result is easy in combinatorics, because the solutions we will be looking for have by construction power series expansion in $\mathbf{t}$\footnote{If we drop the assumption that solutions of \eqref{solsa} must have power series expansion in $\mathbf{t}$, the question of unicity can be addressed under an assumption of strict convexity, see \cite[Section 3]{BEO} in the case where $T_k^0 \equiv 0$ for $k \geq 3$.}.
\begin{lemma}
\label{L36} Assume $\Gamma$ is fixed and the weights $t,\mathbf{t}$ are completely regular. Let $\varepsilon > 0$ and consider $\Omega_{\varepsilon}$ as in \eqref{Ome}. The only solution $\varphi \in \Hs(\Omega_{\varepsilon})$ to the equation:
\beq
\forall z \in \Omega_{\varepsilon}\cap\iota(\Omega_{\varepsilon}),\qquad \mathcal{S}\varphi(z) + \mathcal{O}\varphi(z) = 0,
\eeq
which has a power series expansion in $\mathbf{t}$, is $\varphi \equiv 0$. The same holds if $\mathcal{O}$ is replaced by $\widetilde{\mathcal{O}}$.
\end{lemma}
\textbf{Proof.} Since $\mathcal{O}$ (or $\widetilde{\mathcal{O}}$) depends linearly on the parameters $t_{m_1,\ldots,m_k}^0$, the leading order $\varphi = \varphi_0 + O(\mathbf{t})$ of a power series solution to $\mathcal{S}\varphi(z) + \mathcal{O}\varphi(z) = 0$ satisfies $\varphi_0(z) + \varphi_0(\iota(z)) = 0$, and we remind $\iota(z) = 1/z$. By assumption, $\varphi_0$ is holomorphic in the exterior of the unit disc, and this equation implies that $\varphi_0$ is holomorphic in $\widehat{\mathbb{C}}\setminus \mathbb{U}$. Hence, if $\varphi$ is holomorphic in an open neighborhood of $\mathbb{U}$, so is $\varphi_0$. Gathering all the information, we see that $\varphi_0$ is a holomorphic $1$-form on the Riemann sphere, thus it vanishes. The same argument shows that $\varphi$ cannot have a non-zero leading order in its power series expansion in $\mathbf{t}$, hence it must vanish identically. \hfill $\Box$

\vspace{0.2cm}

Let us comment on the use of this result. Since a power series in a infinite sequence of variable $t,\mathbf{t}$ is characterized by its specializations where all but a finite number of variables have been sent to $0$, it is enough to study the latter. Lemma~\ref{L3} then tells us that, for any given values for the non-zero weights, there exists a neighborhood of $0$ of values of $t$ so that $t,\mathbf{t}^0$ is admissible, and the solutions we will be looking for then have a power series expansion in $t$ with non-zero radius of convergence. Since there are only a finite number of non-zero weights, they are obviously completely regular in the sense of Definition~\ref{comp}. Thus, we do not lose in generality by taking the detour to set $t,\mathbf{t}^0$ to some real admissible values -- which enables us to use the tools of complex analysis -- in order to say something about formal series.

Although we do not pursue this issue here, it is possible to show that $W_1^0(x) \in \mathbb{C}[[x^{-1}]][\mathbf{t}]]$ is uniquely determined by the solution of functional equation \eqref{RHP1} for completely regular weights, together with the requirement that $W_1^0(x)$ is holomorphic in $\mathbb{C}\setminus\Gamma$, is bounded on $\widehat{\mathbb{C}}\setminus\Gamma$, and behaves like $t/x$ when $x \rightarrow \infty$.

\subsection{Local Cauchy kernel}
\label{S36}
We now turn to the generating series of stuffed maps with topology of a cylinder, which will allow us the representation of any solution of the homogeneous linear equation \eqref{solsa}. Cylinders can be obtained by marking an extra elementary $2$-cell with topology of a disc on a stuffed map with topology of a cylinder. At the level of generating series, this means:
\beq
W_2^0(x_1,x_2) = \Big(\sum_{m \geq 1} \frac{1}{x_2^{m + 1}}\,\frac{\partial}{\partial t_{m}^0}\Big)W_1^0(x_1).
\eeq
Applying the differential operator to the functional relation \eqref{RHP1} yields, for all $x_1$ in the interior of $\Gamma_{t,\mathbf{t}^0}$ and $x_2 \in \mathbb{C}\setminus\Gamma$:
\beq
\label{RHP2} W_2^0(x_1 + {\rm i}0,x_2) + W_2^0(x_1 - {\rm i}0,x_2) + \mathcal{O}_{x_1}W_2^0(x_1,x_2) + \frac{1}{(x_1 - x_2)^2} = 0.
\eeq
This equation will also be derived from the analysis of Schwinger-Dyson equation in Section~\ref{S4}. Since $W_2^0(x_1,x_2)$ is symmetric, it satisfies the same equation with respect to $x_2$. The subscript of the operator $\mathcal{O}$ indicates on which variable it acts. So, we can apply Lemma~\ref{L5} to $W_2^0$, and define:
\beq
\mathcal{W}_2^0(z_1,z_2) = W_2^0(x(z_1),x(z_2))\dd x(z_1)\dd x(z_2)
\eeq
as a symmetric meromorphic $2$-form in $(z_1,z_2) \in \Omega_{\varepsilon}$, and it satisfies:
\beq
\mathcal{S}_{z_1}\mathcal{W}_2^0(z_1,z_2) + \mathcal{O}_{z_1}\mathcal{W}_2^0(z_1,z_2) + \frac{\dd x(z_1)\dd x(z_2)}{(x(z_1) - x(z_2))^2}
\eeq
in the domain of analyticity of the left-hand side. We may also define:
\beq
\omega_2^0(z_1,z_2) = \mathcal{W}_2^0(z_1,z_2) + \frac{\dd x(z_1)\dd x(z_2)}{(x(z_1) - x(z_2))^2},
\eeq
which satisfies:
\beq
\mathcal{S}_{z_1}\omega_2^0(z_1,z_2) + \mathcal{O}_{z_1}\omega_2^0(z_1,z_2) = \frac{\dd x(z_1)\dd x(z_2)}{(x(z_1) - x(z_2))^2}.
\eeq
A computation done in the proof of Proposition 3.8 in \cite{BEO} shows that $\omega_2^0(z_1,z_2)$ has its only singularities at $z_1 = z_2$, and it is a double pole with leading coefficient $1$ and no residues. Let us define the \emph{local Cauchy kernel}:
\beq
G(z_0,z) = -\int^{z} \omega_2^0(z_0,\cdot).
\eeq
According to \cite[Lemma 2.1]{BEO}, it allows the representation of any solution of the homogeneous linear equation \eqref{solsa} in terms of its singular part only, modulo a holomorphic part.
\begin{lemma}
Let $\Omega'$ be an open neighborhood of $\mathbb{U}$ stable under $\iota$, and $\varphi \in \Ms(\Omega')$ be a solution of $\mathcal{S}\varphi(z) + \mathcal{O}\varphi(z) = 0$ with a finite number of poles in $\Omega'$. Then,
\beq
\widetilde{\varphi}(z_0) = \sum_{p \in \Omega'} \Res_{z \rightarrow p} \frac{\Delta_{z} G(z_0,z)}{4}\,\Delta \varphi(z)
\eeq
is such that $\varphi(z_0) - \widetilde{\varphi}(z_0)$ is holomorphic for $z_0 \in \Omega'$.
\end{lemma}
We adapt this result to solve \eqref{solsa} with a non-zero right-hand-side:
\begin{lemma}
\label{Lsu} Let $\Omega'$ be an open neighborhood of $\mathbb{U}$ stable under $\iota$, and $\Omega$ be the union of $\Omega'$ and the exterior of the unit disc in $\widehat{\mathbb{C}}$. Let $\psi \in \Hs(\Omega')$. Assume $\varphi \in \Ms(\Omega)$ satisfies  $\mathcal{S}\varphi(z) + \mathcal{O}\varphi(z) + \psi(z) = 0$ for any $z \in \Omega'$, and has a finite number of poles in $\Omega'$. Then, if $z_0$ lies outside the contour of integrations:
\beq
\widetilde{\varphi}(z_0) = - \frac{1}{2{\rm i}\pi} \oint_{\mathbb{U}} G(z_0,z)\,\frac{\psi(z)}{2} + \sum_{p \in \Omega'} \Res_{z \rightarrow p} \frac{\Delta_{z} G(z_0,z)}{4}\,\Delta_{z}\varphi(z)
\eeq
is such that $\varphi(z_0) - \widetilde{\varphi}(z_0)$ is holomorphic for $z_0 \in \Omega$ and satisfies $\mathcal{S}\varphi(z) + \mathcal{O}\varphi(z) = 0$ for $z \in \Omega'$.
\end{lemma}
\textbf{Proof.} It follows from Lemma~\ref{Lsu} and the fact that $\phi(z) = \frac{\psi(z)}{2} -\frac{1}{2{\rm i}\pi}\oint_{\mathbb{U}} G(z_0,z)\,\frac{\psi(z)}{2}$ is holomorphic in $\Omega$, and satisfies $\mathcal{S}\phi(z) + \mathcal{O}\phi(z) + \psi(z) = 0$ for $z \in \Omega'$. \hfill $\Box$

\vspace{0.2cm}

And, if we are looking for a solution $\varphi$ which is initially holomorphic in $\widehat{\mathbb{C}}\setminus\mathbb{U}$, and has a power series expansion in the parameters $\mathbf{t}^0$ of $\mathcal{O}$, we deduce from Lemma~\ref{L36} that $\widetilde{\varphi}(z_0) = \varphi(z_0)$.

\section{Schwinger-Dyson equations and consequences}
\label{S4}

\subsection{Relations between generating series for all topologies}

Stuffed maps $\mathcal{M}$ of genus $g$ with $1$ boundary $\mathfrak{f}$ can be constructed recursively by Tutte's decomposition. It consists in removing the root edge of the first boundary, and establishing a bijection between the set of stuffed maps with given topology, and the pieces obtained after the removal. According to their topology, several cases can occur:
\begin{itemize}
\item[$\bullet$] the root edge $\mathfrak{e}$ was bordered on both sides by $\mathfrak{f}$, and its removal disconnects the surface. We obtain two connected stuffed maps $\mathcal{M}_1$ and $\mathcal{M}_2$, each having one boundary coming from the splitting of $\mathfrak{f}$, and which are rooted at the edge which was closest to $\mathfrak{e}$ following $\mathfrak{f}$ in cyclic order. The handles of $\mathcal{M}$ are shared between $\mathcal{M}_1$ and $\mathcal{M}_2$.
\item[$\bullet$] the root edge borders another elementary $2$-cell $\mathfrak{f}'$ with $k \geq 1$ boundaries. We denote $K = \ldbrack 1,k \rdbrack$ the set of boundaries. Removing the root edge also removes $\mathfrak{f}'$, and we obtain a stuffed map with $r \leq k$ connected components, $\mathcal{M}_{1},\ldots,\mathcal{M}_{r}$. $\mathcal{M}_i$ has $f_i$ handles and $k_i \geq 1$ boundaries, which were incident to a subset $K_i \subseteq K$ of $|K_i| = k_i$ boundaries of $\mathfrak{f}'$. One of the boundary in $\mathcal{M}_1$ was incident to $\mathfrak{f}$ in $\mathcal{M}$. The gluing of $\mathcal{M}_i$ on $\mathfrak{f}'$ contributed to $k_i - 1 + f_i$ handles in $\mathcal{M}$, and $\mathfrak{f}'$ itself contributed for $h$ handles. Therefore, we must have $h + \sum_{i = 1}^r (k_i - 1 + f_i) = g $, which can be rewritten $h + \big(\sum_{i = 1}^r f_i\big)  + k - r = g$.
\end{itemize}

In terms of generating series, this bijection implies, for $g = 0$:
\beq
\label{4} \big(W_1^0(x)\big)^2 + \sum_{k \geq 1} \oint \prod_{j = 1}^k \frac{\dd\xi_j}{2{\rm i}\pi}\,\frac{\partial_{\xi_i} T_k^0(\xi_1,\ldots,\xi_k)}{(k - 1)!\,(x - \xi_1)}\,\prod_{j = 1}^k W_{1}^0(\xi_j) = 0,
\eeq
In this equation, the contour integral is just a way to write the divergent part of a formal Laurent series:
\beq
\oint \frac{\dd\xi}{2{\rm i}\pi}\,\frac{1}{x - \xi}\,\Big(\sum_{m \geq m_0} \frac{\beta_m}{\xi^{m + 1}}\Big) = \sum_{m \geq 0} \frac{\beta_m}{x^{m + 1}}.
\eeq
It enforces the matching of perimeters when reconstructing $\mathcal{M}$ from its pieces after Tutte's decomposition. In this formal representation, everything happens as if the contour was surrounding $\infty$ and $x$ was closer to $\infty$ than the contour. Similarly, for $g > 0$:
\bea
W_2^{g - 1}(x,x) + \sum_{f = 0}^{g} W_1^{f}(x,x_J)W_{1}^{g - f}(x,x_{I\setminus J}) && \nonumber \\
\label{3} + \sum_{\substack{k \geq 1 \\ h \geq 0}}\sum_{\substack{K \vdash \ldbrack 1,k \rdbrack \\ f_1,\ldots,f_{[K]} \geq 0 \\ h + (\sum_i f_i) + k - [K] =  g}} \oint \Big[\prod_{j = 1}^k \frac{\dd\xi_j}{2{\rm i}\pi}\Big]\frac{\partial_{\xi_1} T_k^h(\xi_1,\ldots,\xi_k)}{(k - 1)!\,(x - \xi_1)}\,\prod_{i = 1}^{[K]} W_{|K_i|}^{f_i}(\xi_{K_i}) & = & 0.
\eea
Those relations are equalities between formal series in $\mathbb{C}[[x^{-1}]][[\mathbf{t}]]$, and \eqref{3} is still valid for $g = 0$ with the convention that $W_n^{g} = 0$ if $g < 0$.

To obtain relations for stuffed maps of genus $g$ with an arbitrary number $n \geq 1$ of boundaries, we apply the operator $\delta_{x_2}\cdots\delta_{x_n}$ to \eqref{3}, since
\beq
\label{insert}\delta_{x} = \sum_{m \geq 1} \frac{1}{x^{m + 1}}\,\frac{\partial}{\partial t^0_m}
\eeq
amounts to mark an elementary $2$-cell with topology of a disc, with the formal variable $x$ coupled to its perimeter. $\delta_{x}$ is called the \emph{insertion operator}. The result is, for any $n \geq 1$ and $g \geq 0$:
\bea
\label{Wng}W_{n + 1}^{g - 1}(x,x,x_I) + \sum_{J \subseteq I,\,\,0 \leq f \leq g} W_{|J| + 1}^{f}(x,x_J)W_{n - |J|}^{g - f}(x,x_{I\setminus J}) & & \\
+ \sum_{i \in I} \partial_{x_i}\Big(\frac{W_{n - 1}^g(x,x_{I\setminus\{i\}}) - W_{n - 1}^g(x_I)}{x - x_i}\Big) & & \nonumber \\ 
+ \sum_{\substack{k \geq 1 \\ h \geq 0}}  \sum_{\substack{K \vdash \ldbrack 1,k \rdbrack \\ J_1\dot{\cup}\cdots \dot{\cup} J_{[K]} = I}} \sum_{\substack{f_1,\ldots,f_{[K]} \geq 0 \\ h + (\sum_i f_i) + k - [K] = g}} \oint \Big[\prod_{j = 1}^k \frac{\dd\xi_j}{2{\rm i}\pi}\Big] \frac{\partial_{\xi_1} T_k^{h}(\xi_1,\ldots,\xi_k)}{(k - 1)!\,(x - \xi_1)} \,\prod_{i = 1}^{[K]} W_{|K_i| + |J_i|}^{f_i}(\xi_{K_i},x_{J_i}) & = & 0. \nonumber
\eea
If $W_n^g$ can be upgraded to holomorphic functions of $x_i$ in some domain of the complex plane, \eqref{Wng} will hold in the whole domain of analyticity.

We can rewrite those equations in a more compact way by summing over genera with weight $(N/t)^{\chi}$ and recalling the definitions \eqref{212}-\eqref{213}. Introducing:
\beq
T_k(x_1,\ldots,x_k) = \sum_{h \geq 0} (N/t)^{2 - 2h - k}\,T_k^h(x_1,\ldots,x_k),
\eeq
we find:
\bea
\label{bu} W_{n + 1}(x,x,x_I) + \sum_{J \subseteq I} W_{|J| + 1}(x,x_J)W_{n - |J|}(x,x_{I\setminus J}) && \\
+ \sum_{i \in I} \partial_{x_i}\Big(\frac{W_{n - 1}(x,x_{I\setminus\{i\}}) - W_{n - 1}(x_I)}{x - x_i}\Big) & & \nonumber \\
+ \sum_{k \geq 1} \sum_{\substack{K \vdash \ldbrack 1,k \rdbrack \\ J_1\dot{\cup} \cdots \dot{\cup} J_{[K]} = I}} \oint \Big[\prod_{j = 1}^k \frac{\dd\xi_j}{2{\rm i}\pi}\Big]\frac{\partial_{\xi_1} T_k(\xi_1,\ldots,\xi_k)}{(k - 1)!\,(x - \xi_1)}\,\prod_{i = 1}^{[K]} W_{|K_i| + |J_i|}(\xi_{K_i},x_{J_i}) & = & 0. \nonumber 
\eea
The equations \eqref{bu} can also be derived by integration by parts in the matrix integrals described in \eqref{S21}, or by expressing the invariance of the matrix integral under infinitesimal change of variables $M \rightarrow M + \frac{\varepsilon}{x - M}$. In this context, they are called Schwinger-Dyson equations, and they also hold for convergent integrals.

\subsection{Analytical properties}
\label{S42}
\begin{definition}
\label{tamde} We say that $t,\mathbf{t}$ is tame if $t,\mathbf{t}^0$ is completely regular (see Definition~\ref{comp}), and if for any $m \geq 1$, $h \geq 0$, any partition $M \vdash \ldbrack 1,m \rdbrack$, any sequence $(f_i)_{1 \leq i \leq [M]}$ of nonnegative integers, any finite set $I$, any sequence $(J_i)_{1 \leq i \leq [M]}$ of pairwise disjoint and maybe empty subsets whose union is $I$, the formal series
\beq
\label{jupe}\sum_{k \geq m} \oint \Big[\prod_{j  = 1}^k \frac{\dd\xi_j}{2{\rm i}\pi}\Big] \frac{\partial_{x} T_k^h(x,\xi_2\ldots,\xi_k) - \partial_{\xi_1} T_k^h(\xi_1,\ldots,\xi_k)}{x - \xi_1} \prod_{i = 1}^{[M]} W_{|M_i| + |J_i|}^{f_i}(\xi_{M_i},x_{J_i})\prod_{j = m + 1}^k W_1^0(\xi_j)
\eeq
which belongs a priori to $\mathbb{C}[[x,(x_i^{-1})_{i \in I}]][[t,\mathbf{t}]]$, is a holomorphic function of $x$ in a neighborhood of $\Gamma_{t,\mathbf{t}^0}$ and $x_i$ in a neighborhood of $\infty$.
\end{definition}

Although technical, this condition is similar for usual maps to asking that the model be not critical. It is thus slightly stronger than asking that the coefficients of the generating series considered are finite. This condition allows conveniently the use of analytic functions instead of formal series.  If for any $h \geq 1$, the number of boundaries of elementary $2$-cells and their perimeter are bounded (i.e. only a finite number of $t^h_{\ell_1,\ldots,\ell_k}$ are non-zero for a given $h$). As we already said in \S~\ref{S35}, since a formal series in an infinite number of variables are determined by all their restrictions to finitely many variables, the analytic study we are going to do within the tame condition actually determines completely the generating series of stuffed maps as a formal series in the infinite set of variables $t,\mathbf{t}$.
 
In this paragraph, we upgrade that the generating series of stuffed maps to analytic functions, and study their basic properties.
   
\begin{lemma}
\label{lpole} Assume $t,\mathbf{t}$ is tame, then $W_k^g(x_1,\ldots,x_k)$ defines a holomorphic function in $\mathbb{C}\setminus\Gamma_{t,\mathbf{t}^0}$, which have boundaries values when $x_i$ approaches an interior point of $\Gamma_{t,\mathbf{t}^0}$, and for any $\alpha = a,b$, there exists an integer $r_{\alpha,k}^{g}$ so that $(x_i - \alpha)^{r_{\alpha,k}^g}W_k^g(x_1,\ldots,x_k)$ remains bounded when $x_i \rightarrow \alpha$.
\end{lemma}

\noindent \textbf{Proof.} The statement was established for $(n,g) = (1,0)$ in Lemma~\ref{L1}. Let $(n,g) \neq (1,0)$, and assume the result is proved for $(n',g')$ such that $2g' - 2 + n' < 2g - 2 + n$. We introduce:
\beq
\label{Pkhs} P_k^h(x,\xi_1;\xi_2,\ldots,\xi_k) = \frac{\partial_{x} T_k^h(x,\xi_2,\ldots,\xi_k) - \partial_{\xi_1}T_k^h(\xi_1,\ldots,\xi_k)}{x - \xi_1}.
\eeq
For any $(n,g) \neq (1,0)$, we isolate the contribution of $W_n^g$ in \eqref{Wng} and decompose:
\bea
\label{ua}\big(2W_1^0(x) + \widetilde{\mathcal{O}}W_1^0(x) + \partial_{x}T_1^0(x)\big)W_n^g(x,x_I) + W_1^0(x)\,\mathcal{O}_{x}W_n^g(x,x_I) &&  \\
- \sum_{k \geq 1} \oint \dd\xi_1\,\frac{P_{k}^{0}(x,\xi_1;\xi_2,\ldots,\xi_k)}{(k - 1)!}\,W_n^g(\xi_1,x_I)\Big[\prod_{j = 2}^k \frac{W_1^0(\xi_j)\dd\xi_j}{2{\rm i}\pi}\Big]  && \nonumber \\
- \sum_{k \geq 1} \oint \dd\xi_1\,\frac{P_{k}^0(x,\xi_1;\xi_2,\ldots,\xi_k)}{(k - 2)!}\,W_n^g(\xi_2,x_I)\Big[\prod_{\substack{j = 1 \\ j \neq 2}}^k \frac{W_1^0(\xi_j)\dd\xi_j}{2{\rm i}\pi}\Big] && \nonumber \\ 
+ W_{n + 1}^{g - 1}(x,x,x_I) + \sum_{J \subseteq I,\,\,0 \leq f \leq g}' W_{|J| + 1}^{f}(x,x_J)W_{n - |J|}^{g - f}(x,x_{I\setminus J}) & & \nonumber \\
+ \sum_{i \in I} \partial_{x_i}\Big(\frac{W_{n - 1}^{g}(x,x_{I\setminus\{i\}}) - W_{n - 1}^{g}(x_I)}{x - x_i}\Big) & & \nonumber \\ 
+ \sum_{\substack{k \geq 1 \\ h \geq 0}} \sum_{\substack{K \vdash \ldbrack 1,k \rdbrack \\ J_1\dot{\cup}\cdots \dot{\cup} J_{[K]} = I \\ \xi_1 = x}} \sum_{\substack{f_1,\ldots,f_{[K]} \geq 0 \\ h + (\sum_i f_i) + k - [K] = g}}^{'}  \oint \Big[\prod_{j = 2}^k \frac{\dd\xi_j}{2{\rm i}\pi}\Big] \frac{\partial_{x} T_k^{h}(x,\xi_2,\ldots,\xi_k)}{(k - 1)!}\,\prod_{i = 1}^{[K]} W_{|K_i| + |J_i|}^{f_i}(\xi_{K_i},x_{J_i}) & & \nonumber \\
- \sum_{\substack{k \geq 1 \\ h \geq 0}} \sum_{\substack{K \vdash \ldbrack 1,k \rdbrack \\ J_1\dot{\cup}\cdots \dot{\cup} J_{[K]} = I}} \sum_{\substack{f_1,\ldots,f_{[K]} \geq 0 \\ h + (\sum_i f_i) + k - [K] = g}}'  \oint \Big[\prod_{j = 1}^k \frac{\dd\xi_j}{2{\rm i}\pi}\Big] \frac{P_{k}^h(x,\xi_1;\xi_2,\ldots,\xi_k)}{(k - 1)!} \,\prod_{i = 1}^{[K]} W_{|K_i| + |J_i|}^{f_i}(\xi_{K_i},x_{J_i}) & = & 0, \nonumber
\eea
where $\sum'$ means that we excluded all terms containing $W_n^g$. We see that \eqref{ua} involves only a finite number of terms of the form \eqref{jupe}, and assuming $t,\mathbf{t}$ tame actually justifies the existence of the decomposition \eqref{ua}, and implies that the second, third and last line of \eqref{ua} define holomorphic functions of $x$ in a neighborhood of $\Gamma_{t,\mathbf{t}^0}$. Then, we can write:
\beq
W_n^g(x,x_I) = \frac{L_{n}^g(x;x_I)}{2W_1^0(x) + \mathcal{O}W_1^0(x) + \partial_{x}T_1^0(x)}.
\eeq
We now come to the key observation. $L_n^g(x;x_I)$ involve terms which either:
\begin{itemize}
\item[$\bullet$] define holomorphic functions in an open neighborhood of $\Gamma_{t,\mathbf{t}^0}$. This is the case for $\mathcal{O}_{x}W_n^g(x,x_I)$ and the lines involving the $P$'s.
\item[$\bullet$] or define holomorphic functions in $\mathbb{C}\setminus\Gamma_{t,\mathbf{t}^0}$, since they involve only $W_{n'}^{g'}$ with $2g' - 2 + n' < 2g - 2 + n$ for which we already have the induction hypothesis. 
\end{itemize}
Therefore, $W_n^g(x,x_I)$ upgrades to a holomorphic function in $\mathbb{C}\setminus\Gamma_{t,\mathbf{t}^0}$, and \eqref{ua} is valid in the whole domain of analyticity. From the two points above, we infer that $L_n^g(x)$ behaves as $O((x - \alpha)^{-s_{n}^g/2})$ for some integer $s_{n}^g$ when $x \rightarrow \alpha = a,b$, and has boundary values at any interior point of $\Gamma_{t,\mathbf{t}^0}$. Furthermore, $2W_1^0(x) + \widetilde{\mathcal{O}}W_1^0(x) + \partial_{x}T_1^0(x)$ vanishes like $O(\sqrt{x - \alpha})$ when $x \rightarrow \alpha$, and does not vanish elsewhere on $\Gamma_{t,\mathbf{t}^0}$. Thus $W_n^g(x,x_I) \in O\big((x - \alpha)^{-(s_n^g + 1)/2}\big)$ when $x \rightarrow \alpha$. We thus conclude the proof by induction. $\Box$ \hfill

\subsection{Potentials for higher topologies} 
\label{potS}
In this section, we introduce and study generating series called \emph{potentials for topology $(n,g)$}:
\beq
V_n^{g}(x;x_2,\ldots,x_n) \in \mathbb{C}[[x,x_2^{-1},\ldots,x_n^{-1}]][[\mathbf{t}]],
\eeq
which will appear in the determination of the monodromy of $W_n^g$'s around their discontinuity locus. The cases $(n,g) = (1,0)$ and $(2,0)$ have a special definition:
\beq
V_1^0(x) = T_1^0(x),\qquad V_2^0(x;x_2) = -\frac{1}{x - x_2}.
\eeq
$T_1^0(x)$ is the potential in the usual sense in random matrix theory, and here in the context of multi-trace matrix models, we may call it "potential for discs". For any $(n,g) \neq (1,0),(2,0)$, denoting $I$ a set with $n - 1$ elements, we define the potential in topology $(n,g)$ by:
\bea
\label{Vng} V_n^g(x;x_I) & = & \sum_{\substack{m \geq 1 \\ k \geq m + 1 \\ h \geq 0}} \!\!\sum_{\substack{M \vdash \ldbrack 1,m \rdbrack \\ f_1,\ldots,f_{[M]} \geq 0 \\ h + (\sum_i f_i) + m - [M] = g \\ J_1 \dot{\cup}\cdots \dot{\cup} J_{[M]} = I}}^{'} \oint \Big[\prod_{j = 2}^{k} \frac{\dd\xi_j}{2{\rm i}\pi}\Big] \nonumber \\
&&\qquad\qquad  \times\Big(\frac{m!\,T_k^h(x,\xi_1,\ldots,\xi_{k - 1})}{(k - 1 - m)!} \prod_{i = 1}^{[M]} W_{|M_i| + |J_i|}^{f_i}(\xi_{M_i},x_{J_i}) \prod_{j = m + 1}^{k - 1} W_1^0(\xi_j)\Big). \nonumber
\eea
The $\sum'$ means that we exclude the term which contains $W_n^{g}$, which is actually equal to \mbox{$\mathcal{O}_{x}W_n^{g}(x,x_I)$}. Notice that the variables $x_2,\ldots,x_n$ play symmetric roles, whereas $x$ plays a special role. Besides, those potentials for $2g - 2 + n > 0$ depends on the data $T_k^h$ of generating series of elementary $2$-cells which define the model, but also on the generating series of stuffed maps themselves. Yet, the potential for topology $(n,g)$ only involves the generating series of stuffed maps $W_{n'}^{g'}$ with lower topology, i.e. $2g' - 2 + n' < 2g - 2 + n$.

Combinatorially, $V_n^g(x;x_2,\ldots,x_n)$ is the generating series of one elementary $2$-cell of arbitrary topology $(k,h)$, whose first boundary is unrooted and has a perimeter coupled to $x$, and whose $(k - 1)$ other boundaries are glued to the boundaries of other stuffed maps, so as to form a connected stuffed map $\mathcal{M}$ of genus $g$ with $n$ boundaries, and with the restriction that no stuffed map of topology $(n,g)$ should be used. More precisely, the first boundary of $\mathcal{M}$ is the distinguished boundary of the elementary $2$-cell, while the other boundaries are rooted and their perimeters are coupled to the variables $x_2,\ldots,x_n$. We may describe $\mathcal{M}$ as a stuffed elementary $2$-cell of topology $(n,g)$.

An equivalent way to write the sum in \eqref{Vng} is:
\beq
\label{Vng2} \boxed{V_n^g(x;x_I) = \sum_{\substack{k \geq 2 \\ h \geq 0}} \sum_{\substack{K \vdash \ldbrack 2,k \rdbrack \\ f_1,\ldots,f_{[K]} \geq 0 \\ h + (\sum_i f_i) + k - |K| = g \\ J_1 \dot{\cup}\cdots \dot{\cup} J_{[K]} = I}}^{'} \oint_{\Gamma^{k - 1}} \Big[\prod_{j = 2}^k \frac{\dd\xi_j}{2{\rm i}\pi}\Big]\,\frac{T_k^h(x,\xi_2,\ldots,\xi_k)}{(k - 1)!} \prod_{i = 1}^{[K]} W_{|K_i| + |J_i|}^{f_i}(\xi_{K_i},x_{J_i}).}
\eeq
If $t,\mathbf{t}$ is tame in the sense of Definition~\ref{tamde}, one can deduce that \eqref{Vng2} defines a holomorphic function of $x$ in a neighborhood of $\Gamma_{t,\mathbf{t}^0}$.

$V_n^g(x;x_I)$ can be obtained from $V_1^g(x)$ by successive applications of the insertion operators \eqref{insert} $\delta_{x_i}$ for $i \in I$, since we have the relation:
\beq
\label{Vngrec} \delta_{y}\big(\mathcal{O}_{x}W_n^g(x,x_I) + V_{n}^g(x;x_I)\big) = \mathcal{O}_{x}W_{n + 1}^{g}(x,y,x_I) + V_{n + 1}^{g}(x;y,x_I).
\eeq
For later use, we give a formula for $(n + 1,g - 1) \neq (1,0),(2,0)$:
\bea
\label{Vng3}\!\!\!\!\!\!&& \partial_{1} V_{n + 1}^{g - 1}(x,x,x_I)  \\
\!\!\!\!\!\!&& = \lim_{y \rightarrow x} \partial_{x} V_{n + 1}^{g - 1}(x,y,x_I) \nonumber \\
\!\!\!\!\!\! && = \sum_{\substack{k \geq 2 \\ h \geq 0}} \!\!\!\!\sum_{\substack{K \vdash \ldbrack 2,k \rdbrack \\ f_1,\ldots,f_{[K]} \geq 0 \\ h + (\sum_i f_i) + k - [K] = g - 1 \\ J_1\dot{\cup} \cdots \dot{\cup} J_{[K]} = I}}^{'} \!\!\!\!\!\!\!\!\!\!\!\!\!\oint \Big[\prod_{j = 2}^{k} \frac{\dd\xi_j}{2{\rm i}\pi}\Big]\,\frac{\partial_x T_k^h(x,\xi_2,\ldots,\xi_k)}{(k - 1)!} W_{|K_1| + |J_1| + 1}^{f_1}(x,\xi_{K_1},x_{J_1}) \prod_{i = 2}^{[K]} W_{|K_i| + |J_i|}^{f_i}(\xi_{K_i},x_{J_i}). \nonumber
\eea
It is readily checked from \eqref{Vng2} by calling $1$ the index of the element of the partition $K$ for which the corresponding subset of $I\cup\{y\}$ in the contains the variable $y$.

\subsection{Monodromy of $W_n^g$'s}
\label{S44}
We establish the analog of \eqref{RHP1} for generating series of stuffed maps of higher topologies:
\begin{theorem}
\label{thet}For any $x$ interior to $\Gamma$, and any $x_2,\ldots,x_n \in \mathbb{C}\setminus\Gamma$, we have:
\beq
\label{result}\mathcal{S}_{x}W_n^g(x,x_I) + \mathcal{O}_{x}W_n^g(x,x_I) + \partial_{x} V_{n}^g(x,x_I) = 0,
\eeq
where $V_n^g$ is the potential for topology $(n,g)$ introduced in \eqref{Vng2}.
\end{theorem}
As a consequence of Lemma~\ref{lpole} and Theorem~\ref{thet}, following \S~\ref{S34}, there exists a symmetric $n$-form in $n$ variables $\mathcal{W}_n^g(z_1,\ldots,z_n)$, holomorphic when $z_1,\ldots,z_n$ belong to the exterior of $\mathbb{U}$ in $\mathbb{C}$ and such that:
\beq
\label{419}\mathcal{W}_{n}^g(z_1,\ldots,z_n) = W_n^g(x(z_1),\ldots,x(z_n))\dd x(z_1)\cdots \dd x(z_n),
\eeq
and meromorphic when one of the $z_i$ is in a neighborhood of $\mathbb{U}$. Similarly, we have a function of $z$:
\beq
\mathcal{V}_n^g(z;z_I) = V_n^g(x(z),x(z_I)) \prod_{i \in I} \dd x(z_i)
\eeq
which is holomorphic when $z$ is in a neighborhood of $\mathbb{U}$ stable under $\iota$, and such that \mbox{$\mathcal{V}_n^g(\iota(z);z_I) = \mathcal{V}_n^g(z;z_I)$} in this neighborhood.
Besides, if $z_I$ is a set of $(n - 1)$ spectator variables in the domain of analyticity, and $z \in \Omega_{\varepsilon} \cap \iota(\Omega_{\varepsilon})$ for some $\varepsilon > 0$, \eqref{result} translates into:
\beq
\boxed{\label{result2}\mathcal{S}_{z}\mathcal{W}_n^g(z,z_I) + \mathcal{O}_{z}\mathcal{W}_{n}^g(z,z_I) + \dd_{z}\mathcal{V}_n^g(z;z_I) = 0}
\eeq
The definition of $\mathcal{W}_1^0(z)$ and $\mathcal{W}_2^0(z_1,z_2)$, as well as their analytic properties, were already treated in \S~\ref{S34}-\ref{S36}.

\vspace{0.2cm}

\noindent\textbf{Proof.} We recall the definitions:
\beq
\mathcal{S}\phi(x) = \phi(x + {\rm i}0) + \phi(x - {\rm i}0),\qquad \Delta\phi(x) = \phi(x + {\rm i}0) - \phi(x - {\rm i}0).
\eeq
We have the polarization formulas:
\bea
\mathcal{S}(\phi_1\cdot\phi_2)(x) & = & \frac{1}{2}\big(\mathcal{S}\phi_1(x)\cdot\mathcal{S}\phi_2(x) + \Delta \phi_1(x)\cdot\Delta\phi_2(x)\big), \nonumber \\
\Delta(\phi_1\cdot\phi_2)(x) & = & \frac{1}{2}\big(\mathcal{S}\phi_1(x)\cdot\Delta\phi_2(x) + \Delta\phi_1(x)\cdot\mathcal{S}\phi_2(x)\big).
\eea
We will compute the discontinuity of the Schwinger-Dyson equations in the form \eqref{ua}, and we remind that the terms involving $\mathcal{O}\phi(x), \widetilde{\mathcal{O}}\phi(x)$ and the $P$'s are holomorphic in a neighborhood of $\Gamma$, thus have no discontinuity across $\Gamma$. For $g = 0$, there is a huge simplification in the sum over partitions $K \vdash \ldbrack 1,k \rdbrack$, since we must have $h + \big(\sum_i f_i\big) + k - [K] = 0$, therefore $h = f_1 = \ldots = f_{[K]} = 0$ and $[K] = k$, which means that $K$ is the partition consisting of singletons. We will consider the cases $(n,g) = (1,0),(2,0),(1,1)$ which are somewhat special, before explaining the general pattern of the proof, which proceeds by induction on $2g - 2 + n$. It is possible to derive the result for all $(n,g)$ from the result for all $(n = 1,g)$ by successive applications of the insertion operator using \eqref{Vngrec} (one should not forget to act with $\delta_{x}$ on the operator $\mathcal{O}$). We will take a more direct route, which has its own pedagogical interest, although it is more cumbersome.

For $(n,g) = (1,0)$, the Schwinger-Dyson equation only involves $W_1^0$. Therefore, the $\sum'$ are empty, and we easily find:
\beq
\Delta_{x}\big[\big(W_1^0(x)\big)^2\big] + \Delta_{x}W_1^0(x)\big(\widetilde{\mathcal{O}}_{x}W_1^0(x) + \partial_{x}T_1^0\big) = 0.
\eeq
Using the polarization formula to transform the first term, we infer:
\beq
\Delta_{x}W_1^0(x)\big(\mathcal{S}_{x}W_1^0(x) + \widetilde{\mathcal{O}}_{x}W_1^0(x) + \partial_x T_1^0(x)\big) = 0.
\eeq
Hence, we retrieve the equation \eqref{RHP1} stating that, on the discontinuity locus (the interior of $\Gamma$) of $W_1^0$:
\beq
\label{cut10}\mathcal{S}_{x}W_1^0(x) + \mathcal{O}_{x}W_1^0(x) + \partial_x T_1^0(x) = 0.
\eeq
By definition $V_1^0(x) = T_1^0(x)$, hence \eqref{result} for $(n,g) = (1,0)$.

For $(n,g) = (2,0)$, the set indexing auxiliary variables is $I = \{2\}$, hence in the sum over $(J_i)_{1 \leq i \leq k}$ in the Schwinger-Dyson equation \eqref{Wng}, we just have to choose in which $J_i$ we put the element $2$. We get  first term if we put $2$ in $J_1$, and $(k - 1)$ equal terms for $2 \notin J_1$. So, the Schwinger-Dyson equation reads:
\bea
2W_1^0(x)W_2^0(x,x_2) + \partial_{x_2}\Big(\frac{W_1^0(x) - W_1^0(x_2)}{x - x_2}\Big) && \\
+ \sum_{k \geq 1} \oint_{\Gamma^{k}} \Big[\prod_{j = 1}^k \frac{\dd\xi_j}{2{\rm i}\pi}\Big] \frac{\partial_{\xi_1} T_k^0(\xi_1,\ldots,\xi_k)}{(k - 1)!\,(x - \xi_1)}\,W_2^0(\xi_1,x_2)\prod_{i = 2}^{k} W_1^0(\xi_i) && \\
+ \sum_{k \geq 2} \oint_{\Gamma^{k}} \Big[\prod_{j = 1}^k \frac{\dd\xi_j}{2{\rm i}\pi}\Big]\frac{\partial_{\xi_1} T_k^0(\xi_1,\ldots,\xi_k)}{(k - 2)!\,(x - \xi_1)}\,W_1^0(\xi_1)W_2^0(\xi_2,x_2)\prod_{i = 3}^{k} W_1^0(\xi_i) & = & 0.
\eea
Then, computing its discontinuity with respect to $x$ and applying the polarization formula for the first term, we find:
\bea
\Delta_{x}W_1^0(x)\,\mathcal{S}_xW_2^0(x,x_2) + \mathcal{S}_{x}W_1^0(x)\,\Delta_{x} W_2^0(x,x_2) + \frac{1}{(x - x_2)^2} && \nonumber \\
+ \Delta_{x}W_1^0(x)\,\mathcal{O}_{x}W_2^0(x,x_2) + \Delta_{x}W_2^0(x,x_2)\big(\widetilde{\mathcal{O}}_{x}W_1^0(x) + \partial_x T_1^0(x)\big) & = & 0. \nonumber
\eea
We collect the terms:
\bea
\Delta_{x}W_1^0(x)\Big(\mathcal{S}_{x}W_2^0(x,x_2) + \mathcal{O}_{x}W_2^0(x,x_2) + \frac{1}{(x - x_2)^2}\Big) & & \nonumber \\
+ \Delta_{x}W_2^0(x,x_2)\big(\mathcal{S}_xW_1^0(x) + \widetilde{\mathcal{O}}_{x}W_1^0(x) + \partial_xT_1^0(x)\big) & = & 0,
\eea
and since $W_1^0$ satisfies \eqref{cut10}, we find for any interior point $x$ of $\Gamma$:
\beq
\label{cut02}\mathcal{S}_xW_2^0(x,x_2) + \mathcal{O}_xW_2^0(x,x_2) + \frac{1}{(x - x_2)^2} = 0.
\eeq
This equation was already derived in \S~\ref{S36} by application of the insertion operator on \eqref{cut10}. Since by definition, $V_2^0(x,x_2) = -\frac{1}{x - x_2}$, we obtain \eqref{result} for $(n,g) = (2,0)$.
 
We now come to $(n,g) = (1,1)$. The Schwinger-Dyson equation \eqref{Wng} reads:
\bea
2W_1^0(x)W_1^1(x) + W_2^0(x,x) && \nonumber \\
+ \sum_{k \geq 1} \oint_{\Gamma^{k}} \Big[\prod_{j = 1}^{k} \frac{\dd\xi_j}{2{\rm i}\pi}\Big] \frac{\partial_{\xi_1} T_k^0(\xi_1,\xi_2,\ldots,\xi_k)}{(k - 1)!\,(x - \xi_1)}\,W_1^1(\xi_1)\prod_{i = 2}^k W_1^0(\xi_i) && \nonumber \\
\label{kiq}  + \sum_{k \geq 2} \oint_{\Gamma^{k}} \Big[\prod_{j = 1}^{k} \frac{\dd\xi_j}{2{\rm i}\pi}\Big] \frac{\partial_{\xi_1} T_k^0(\xi_1,\xi_2,\ldots,\xi_k)}{(k - 2)!\,(x - \xi_1)}\,W_1^0(\xi_1)W_1^1(\xi_2) \prod_{i = 3}^{k} W_1^0(\xi_i) & & \\
+ \sum_{k \geq 2} \oint_{\Gamma^{k}} \Big[\prod_{j = 1}^{k} \frac{\dd\xi_j}{2{\rm i}\pi}\Big] \frac{\partial_{\xi_1} T_k^0(\xi_1,\xi_2,\ldots,\xi_k)}{(k - 2)!\,(x - \xi_1)}\,W_2^0(\xi_1,\xi_2)\prod_{i = 3}^{k} W_1^0(\xi_i) && \\
+ \sum_{k \geq 3} \oint_{\Gamma^{k}} \Big[\prod_{j = 1}^{k} \frac{\dd\xi_j}{2{\rm i}\pi}\Big] \frac{2 \partial_{\xi_1} T_k^0(\xi_1,\xi_2,\ldots,\xi_k)}{(k - 3)!\,(x - \xi_1)}\,W_1^0(\xi_1)W_2^0(\xi_2,\xi_3) \prod_{i = 4}^{k} W_1^0(\xi_i)  && \\
+ \sum_{k \geq 1} \oint_{\Gamma^{k}} \Big[\prod_{j = 1}^{k} \frac{\dd\xi_j}{2{\rm i}\pi}\Big] \frac{\partial_{\xi_1}T_k^1(\xi_1,\xi_2,\ldots,\xi_k)}{(k - 1)!\,(x - \xi_1)} \prod_{i = 1}^k W_1^0(\xi_i) & = & 0.
\eea
The discontinuity of the first term can be computed by polarization formula. For the second term, we write similarly:
\beq
\Delta_{x}\big(W_2^0(x,x)\big) = \lim_{y \rightarrow x} \Delta_{x}\mathcal{S}_{y}W_2^0(x,y).
\eeq
We find for the discontinuity of \eqref{kiq}:
\bea
\Delta_{x}W_1^0(x)\,\mathcal{S}_x W_1^1(x) + \mathcal{S}_{x}W_1^0(x)\,\Delta_{x} W_1^1(x) + \lim_{y \rightarrow x} \Delta_{x}\mathcal{S}_y W_2^0(x,y) && \nonumber \\
+ \Delta_{x}W_1^1(x)\big(\widetilde{\mathcal{O}}W_1^0(x) + \partial_x T_1^0(x)\big) + \Delta_{x}W_1^0(x)\,\mathcal{O}_{x}W_1^0(x) && \nonumber \\
+ \lim_{y \rightarrow x} \Delta_{x}\mathcal{O}_{y}W_2^0(x,y) && \nonumber \\
+ \Delta_{x}W_1^0(x)\Big(\sum_{k \geq 3} \oint_{\Gamma^{k - 1}} \Big[\prod_{j = 2}^{k} \frac{\dd\xi_j}{2{\rm i}\pi}\Big] \frac{2\,\partial_{x} T_k^0(x,\xi_2,\ldots,\xi_k)}{(k - 3)!}W_2^0(\xi_2,\xi_3) \prod_{i = 4}^{k} W_1^0(\xi_i)\Big) && \nonumber \\
\label{rju}+ \Delta_{x}W_1^0(x)\Big(\sum_{k \geq 1} \oint_{\Gamma^{k - 1}} \Big[\prod_{j = 2}^{k} \frac{\dd\xi_j}{2{\rm i}\pi}\Big] \frac{\partial_{x} T_k^1(x,\xi_2,\ldots,\xi_k)}{(k - 1)!} \prod_{i = 2}^k W_1^0(\xi_i)\Big) & = & 0.
\eea
This can be rewritten:
\bea
\lim_{y \rightarrow x} \Delta_{y}\Big(\mathcal{S}_{x}W_2^{0}(x,y) + \mathcal{O}_xW_2^0(x,y) + \frac{1}{(x - y)^2}\Big) && \nonumber \\
 + \big(\mathcal{S}_xW_1^0(x) + \widetilde{\mathcal{O}}_xW_1^0(x) + \partial_x T_1^0(x)\big)\Delta_{x}W_1^1(x) && \\
+ \Delta_{x}W_1^0(x)\big(\mathcal{S}_{x}W_1^1(x) + \mathcal{O}_{x}W_1^1(x) + \partial_x V_{1}^1(x)\big) & = & 0,
\eea
where $V_1^1(x)$ is the potential for tori with one boundary introduced in \eqref{Vng2}, namely:
\bea
V_1^{1}(x) & = & \sum_{k \geq 3} \oint_{\Gamma^{k - 1}} \Big[\prod_{j = 2}^k \frac{\dd\xi_j}{2{\rm i}\pi}\Big]\frac{2\,\partial_x T_k^0(x,\xi_2,\ldots,\xi_k)}{(k - 3)!}\,W_2^0(\xi_2,\xi_3)\prod_{i = 4}^{k} W_1^0(\xi_i) \nonumber \\
& & + \sum_{k \geq 1} \oint_{\Gamma^{k - 1}}\Big[\prod_{j = 2}^k \frac{\dd\xi_j}{2{\rm i}\pi}\Big]\frac{\partial_x T_k^1(x,\xi_2,\ldots,\xi_k)}{(k - 1)!}\,\prod_{i = 2}^k W_1^0(\xi_i).
\eea
In order to obtain \eqref{rju}, we have introduced $\Delta_{y}\big(\frac{1}{(x - y)^2}\big) = 0$ in the equation to recognize the combination appearing in \eqref{cut02}. Since we already have linear equations \eqref{cut10}-\eqref{cut02} for $W_1^0$ and $W_2^0$, we find at any interior point of $\Gamma$:
\beq
\mathcal{S}_{x}W_1^1(x) + \mathcal{O}_{x}W_1^1(x) + \partial_x V_1^1(x) = 0.
\eeq
This case was special in the sense that we had to split $W_2^0(x,x)$ in $\lim_{y \rightarrow x} W_2^0(x,y)$ because of the pole at $x = y$ in the equation \eqref{cut02}. This issue is absent for the other values of $(n,g)$.

We now arrive to the general case. Let $n \geq 1$ and $g \geq 0$ be integers such that $2g - 2 + n > 0$, and $(n,g) \neq (1,1)$. Let us assume that the result \eqref{result} holds for any $W_{n'}^{g'}$ such that $2g' - 2 + n' < 2g - 2 + n$. As before, we compute the discontinuity with respect to $x$ of the Schwinger-Dyson equation \eqref{Wng}. In the sum over partitions $K \vdash \ldbrack 2,k \rdbrack$, we have to distinguish whether the element of $K$ which contained $1$ (associated to the variable $\xi_1$), that we call $K_1$, is a singleton or not. We denote $K'$ the partition of $\ldbrack 1,k \rdbrack \setminus K_1$ determined by the other elements of $K$. We then find:
\bea
\Delta_{x,2}\mathcal{S}_{x,1} W_{n + 1}^{g - 1}(x,x,x_I)  + \Delta_{x}W_1^0(x)\,\mathcal{S}_{x}W_n^g(x,x_I) && \\
+ \sum_{\substack{J \subseteq I,\,\,0 \leq f \leq g \\ (J,f) \neq (\emptyset,0),(I,g)}} \Delta_{x} W_{|J| + 1}^{f}(x,x_J)\,\mathcal{S}_{x} W_{n - |J|}^{g - f}(x,x_{I\setminus J}) + \sum_{i \in I} \frac{\Delta_{x}W_{n - 1}^{g}(x,x_{I\setminus\{i\}})}{(x - x_i)^2} && \nonumber \\
+ \sum_{\substack{k \geq 1 \\ h \geq 0}} \sum_{\substack{J \subseteq I \\ 0 \leq f \leq g}} \sum_{\substack{K' \vdash \ldbrack 2,k \rdbrack  \\ J_1' \dot{\cup}\cdots \dot{\cup} J_{[K']}' = I\setminus J}} \sum_{\substack{f_1',\ldots,f_{[K']}' \geq 0 \\ h + (\sum_{i} f_i') + k - ([K'] + 1) = g - f}} \oint_{\Gamma^{k - 1}} \Big[\prod_{j = 2}^{k} \frac{\dd\xi_j}{2{\rm i}\pi}\Big] \frac{\partial_{x} T_k^h(x,\xi_2,\ldots,\xi_k)}{(k - 1)!} & & \nonumber \\
 \Delta_{x} W_{|J| + 1}^{f}(x,x_J) \prod_{i = 1}^{[K']} W_{|K_i'| + |J_i'|}^{f_i'}(\xi_{K_i'},x_{J_i'}) & & \nonumber \\
+ \sum_{\substack{k \geq 2 \\ h \geq 0}} \sum_{\substack{K' \vdash \ldbrack 2,k \rdbrack \\ f_1,\ldots,f_{[K']} \geq 0 \\ h + (\sum_i f_i) + k - [K'] = g \\ J_1 \dot{\cup} \cdots \dot{\cup} J_{[K']} = I}} \oint_{\Gamma^{k - 1}} \Big[\prod_{j = 2}^k \frac{\dd\xi_j}{2{\rm i}\pi}\Big] \frac{\partial_x T_k^h(x,\xi_2,\ldots,\xi_k)}{(k - 1)!} && \nonumber \\
\Delta_{x}W_{|K_1'| + |J_1| + 1}^{f_1}(x,\xi_{K_1},x_{J_1}) \prod_{i = 2}^{[K']} W_{|K_i'| + |J_i|}^{f_i}(\xi_{K_i'},x_{J_i})
 &= & 0. \nonumber
\eea
The indices $x,i$ for the operators $\Delta$ or $\mathcal{S}$ in the first line indicate on which of the two variable $x$ they act. We can collect the terms in three steps:
\begin{itemize}
\item[$\bullet$] In the second line, $\Delta_x W_{n - 1}^{g}(x,x_{I\setminus\{i\}})/(x - x_i)^2$ can be included in the term $\Delta_{x}W_{n - 1}^g(x,x_{I\setminus\{i\}})\,\mathcal{S}_{x}W_2^0(x,x_i)$ arising in the sum over $J \subseteq I$.
\item[$\bullet$] The prefactor of the terms involving \mbox{$\Delta_{x} W_{|J| + 1}^{f}(x,x_J)$} in the third/fourth line can be included in the term \mbox{$\Delta_{x}W_{|J| + 1}^f(x,x_J)\,\mathcal{S}_{x}W_{n - |J|}^{g - f}(x,x_{I\setminus J})$} of the second line. For $(J,f) \neq (I,g)$, it produces: a term for which $|K_i| + |J_i| = 1$ and $f_i = 0$ for all $i$, which is equal to \mbox{$\Delta_{x}W_{|J| + 1}^f(x,x_j)\,\mathcal{O}_{x}W_{n - |J|}^{g - f}(x,x_{I\setminus J})$} ; and a term equal to \mbox{$\Delta_{x}W_{|J| + 1}^f(x,x_J)\,V_{n - |J|}^{g - f}(x;x_{I \setminus J})$}, by comparison with the definition of the potential for higher topologies \eqref{Vng2}. When $(J,f) = (\emptyset,0)$, the result is slightly different due to symmetry factors, and we obtain a contribution \mbox{$\Delta_{x}W_n^g(x,x_I)\big(\widetilde{\mathcal{O}}W_1^0(x) + \partial_x T_1^0(x)\big)$}.
\item[$\bullet$] The last two lines are equal to $\Delta_{x}V_{n}^g(x;x,x_I)$. This can be checked by comparing the last two lines with the expression of $\partial_{1}V_{n + 1}^{g - 1}(x;x,x_I)$ given in \eqref{Vng3}, and noticing that $\partial_x T_k^h(x,\xi_2,\ldots,\xi_k)$ is by assumption a holomorphic function of $x$ in a neighborhood of $\Gamma$.
\end{itemize}
Therefore, we have found:
\bea
\label{ledto} \Delta_{x}W_1^0(x)\big(\mathcal{S}_{x}W_{n}^g(x,x_I) + \mathcal{O}_{x}W_{n}^g(x,x_I) + \partial_{x} V_n^g(x;x_I)\big) && \\
 \Delta_{x}W_n^g(x,x_I)\big(\mathcal{S}_{x}W_1^0(x) + \widetilde{\mathcal{O}}_{x}W_1^0(x) + \partial_x T_1^0(x)\big) && \nonumber \\
 + \sum_{\substack{J \subseteq I,\,\,0 \leq f \leq g \\ (J,f) \neq (\emptyset,0),(I,g)}} \Delta_{x}W_{|J| + 1}^{f}(x,x_J)\Big(\mathcal{S}_{x} W_{n - |J|}^{g - f}(x,x_{I\setminus J}) + \mathcal{O}_{x}W_{n - |J|}^{g - f}(x,x_{I\setminus J}) + \frac{\delta_{n - |J|,2}\,\delta_{g - f,0}}{(x - x_{I\setminus J})^2}\Big) && \nonumber \\
 + \Delta_{1}\big(\mathcal{S}_{2}W_{n + 1}^{g - 1}(x,x,x_I) + \mathcal{O}_{2}W_{n + 1}^{g - 1}(x,x,x_I) + \partial_{1} V_{n + 1}^{g - 1}(x;x,x_I)\big) & = & 0. \nonumber 
\eea
where the indices on the operators in the last line indicate on which variable $x$ (the first or the second) they act. By the induction hypothesis, the three last lines vanish: we deduce that for any interior point $x$ of $\Gamma$,
\beq
\mathcal{S}_{x}W_n^g(x,x_I) + \mathcal{O}_{x}W_n^g(x,x_I) + \partial_{x} V_n^g(x;x_I) = 0,
\eeq 
which is the desired result.

\hfill $\Box$

\subsection{From Schwinger-Dyson equations to quadratic loop equations}
\label{S45}
We define for convenience
\beq
\label{brevng}\breve{\mathcal{W}}_{n}^g(z_1,\ldots,z_n) = \mathcal{W}_n^g(z_1,\ldots,z_n) + \delta_{n,2}\delta_{g,0}\,\frac{\dd x(z_1)\,\dd x(z_2)}{\big(x(z_1) - x(z_2)\big)^2}
\eeq
The only difference is that now $\breve{\mathcal{W}}_{2}^0(z_1,z_2) = \omega_2^0(z_1,z_2)$, thus has a singularity only at $z_1 = z_2$.
\begin{theorem}
\label{thet2}For any $(n,g) \neq (1,0),(2,0)$, the quadratic differential form in $z$:
\beq
\label{Qde}\mathcal{Q}_n^g(z;z_I) = \breve{\mathcal{W}}_{n + 1}^{g - 1}(z,\iota(z),z_I) + \sum_{\substack{J \subseteq I \\ 0 \leq f \leq g}} \breve{\mathcal{W}}_{|J| + 1}^{f}(z,z_J)\,\breve{\mathcal{W}}_{n - |J|}^{g - f}(\iota(z),z_J)
\eeq
has double zeroes at $z = \pm 1$, i.e. $x(z) \in \{a,b\}$.
\end{theorem}
The content of this theorem is that, although $\mathcal{W}_n^g$ can have poles of high order at $z = \pm 1$, the combination $\mathcal{Q}_n^g(z;z_I)$ does not.

\vspace{0.2cm}

\noindent \textbf{Proof.} To arrive to \eqref{Qde}, we have going to recast the Schwinger-Dyson equation \eqref{ua} using the same decomposition of the sum over partitions $K \vdash \ldbrack 1,k \rdbrack$ which led to \eqref{ledto}. 
We find:
\bea
\label{juikq} \widetilde{\mathcal{Q}}_{n}^{g}(z;z_I) + \mathcal{W}_{n + 1}^{g - 1}(z,z,z_I) + \mathcal{O}_{z,2}\mathcal{W}_{n + 1}^{g - 1}(z,z,z_I) + (1 - \delta_{n,1}\delta_{g,0})\dd_{2}\mathcal{V}_{n + 1}^{g - 1}(z;z,z_I) &&  \\
+ \big(2\mathcal{W}_1^0(z) + \widetilde{\mathcal{O}}_{z}\mathcal{W}_1^0(z) + \dd_{z}\mathcal{V}_1^0(z)\big)\mathcal{W}_n^g(z,z_I) && \nonumber \\
+ \mathcal{W}_1^0(z)\big(\mathcal{O}_{z}\mathcal{W}_n^g(z,z_I) + \dd_{z}\mathcal{V}_n^g(z;z_I)\big) &&  \nonumber \\
+ \sum_{\substack{J \subseteq I,\,\,0 \leq f \leq g \\ (J,f) \neq (\emptyset,0),(I,g)}} \mathcal{W}_{|J| + 1}^{f}(z,z_J)\big(\mathcal{W}_{n - |J|}^{g - f}(z,z_{I\setminus J}) + \mathcal{O}_{z}\mathcal{W}_{n - |J|}^{g - f}(z,z_{I\setminus J}) + \dd_{z}\mathcal{V}_{n - |J|}^{g - f}(z;z_{I\setminus J})\big) & = & 0 \nonumber
\eea
where:
\bea
\label{Qngde} \widetilde{\mathcal{Q}}_n^g(z;z_I) & = & - \dd x(z)\,\dd_{z_i}\Big(\frac{\mathcal{W}_{n - 1}^{g}(z_I)}{\dd x(z_i)\,\big(x(z) - x(z_i)\big)^2}\Big) \\
&- & \sum_{\substack{k \geq 1 \\ h \geq 0}} \sum_{\substack{K \vdash \ldbrack 1,k \rdbrack \\ J_1 \dot{\cup}\cdots \dot{\cup} J_{[K]} = I}} \sum_{\substack{f_1,\ldots,f_{[K]} \geq 0 \\ h + (\sum_i f_i) + k - [K] = g}} \oint_{\mathbb{U}^k} \frac{\mathcal{P}_k^h(z,\zeta_1;\zeta_2,\ldots,\zeta_k)}{(k - 1)!\,\big(x(z) - x(\zeta_1))} \prod_{i = 1}^{[K]} \mathcal{W}_{|K_i| + |J_i|}^{f_i}(\zeta_{K_i},z_{J_i}). \nonumber
\eea
The contribution \mbox{$\sum_{i \in I} \dd x(z)\,\dd_{z_i}\big(\frac{\mathcal{W}_{n - 1}^g(z_I)}{(x(z) - x(z_i))^2}\big)$} to the Schwinger-Dyson equations was included in the term $\mathcal{V}_{2}^0$ appearing in the sum of the fourth line. We have introduced the differential form version of \eqref{Pkhs}, and:
\beq
\mathcal{P}_k^h(z,\zeta_1;\zeta_2,\ldots,\zeta_k) = \frac{\dd x(\zeta_1)\,\dd_{z} T_k^h(x(z),x(\zeta_2),\ldots,x(\zeta_k)) - \dd x(z)\,\dd_{\zeta_1} T_k^h(x(\zeta_1),x(\zeta_2),\ldots,x(\zeta_k))}{x(z) - x(\zeta_1)},
\eeq
and in \eqref{Qngde}, the variables $\zeta_i$ are integrated over the unit circle. We already observe that $\widetilde{\mathcal{Q}}_n^g(z;z_I)$ has a double zero at $z = \{\pm 1\}$, since it is a holomorphic function in a neighborhood of $z = \pm 1$ multiplied by $\big(\dd x(z)\big)^2$. We also recognize in \eqref{juikq} combinations which can be represented using:
\beq
\mathcal{W}_{n'}^{g'}(\iota(z),z_J) = -\mathcal{W}_{n'}^{g'}(z,z_J) - \mathcal{O}_{z}\mathcal{W}_{n'}^{g'}(z,z_J) - \dd_{z}\mathcal{V}_{n'}^{g'}(z,z_J).
\eeq
If we rewrite the equality \eqref{juikq} in terms of $\omega_{n'}^{g'}(z,z_J)$ and $\omega_{n'}^{g'}(\iota(z),z_J)$, we conclude after some algebra that $\mathcal{Q}_{n}^{g}(z;z_I) = \widetilde{\mathcal{Q}}_{n}^g(z;z_I)$. \hfill $\Box$

\section{Solution by the topological recursion}
\label{S5}

\subsection{Main result}
\label{S66}
Assuming that $t,\mathbf{t}$ are tame, we are going to show that the generating series of stuffed maps $W_n^g$ (in the $x$ variables) or $\mathcal{W}_n^g$ (in the $z$ variables), are given up to a shift -- which is essential -- by the topological recursion of \cite{EOFg} applied to the initial data:
\beq
\label{de21}\omega_1^0(z) = \mathcal{W}_1^0(x(z))\dd x(z),
\eeq
together with the Bergman kernel:
\beq
\label{de22}\omega_2^0(z_1,z_2) = \Big(\mathcal{W}_2^0(x(z_1),x(z_2)) + \frac{1}{\big(x(z_1) - x(z_2)\big)^2}\Big)\dd x(z_1)\dd x(z_2),
\eeq
and local involution given by $\iota(z) = 1/z$ (this is Theorem~\ref{tpoth} below).

For this purpose, we remind the definition of the local Cauchy kernel:
\beq
G(z_0,z) = -\int^{z} \omega_2^0(z_0,\cdot),
\eeq
and introduce the \emph{recursion kernel}:
\beq
K(z_0,z) = -\frac{1}{2}\,\frac{\Delta_{z} G(z_0,z)}{\Delta_{z}\omega_1^0(z)} = -\frac{\frac{1}{2}\int_{\iota(z)}^{z} \omega_2^0(z_0,\cdot)}{\omega_1^0(z) - \omega_1^0(\iota(z))}.
\eeq
For $2g - 2 + n > 0$, we introduce the meromorphic forms:
\beq
\omega_{n}^g(z_1,\ldots,z_n) = \mathcal{W}_n^g(z_1,\ldots,z_n)\dd x(z_1)\cdots\dd x(z_n)
\eeq
and from Theorem~\ref{thet}, we have the inhomogeneous linear equations:
\beq
\label{thet3}\mathcal{S}_{z}\omega_n^g(z,z_I) + \mathcal{O}_{z}\omega_{n}^g(z,z_I) + \dd_{z}\mathcal{V}_{n}^g(z,z_I) = 0
\eeq
According to Lemma~\ref{L5} and Lemma~\ref{lpole}, $\omega_n^g(z,z_I)$ is meromorphic in a neighborhood of $\mathbb{U}$, and has poles only at $z = \pm 1$. Therefore, and since we are working in the realm of formal series in $t,\mathbf{t}$, we can apply Lemma~\ref{Lsu} and Lemma~\ref{L36} to represent, for $(n,g) \neq (1,0),(2,0)$:
\bea
\label{repsa}\omega_n^g(z,z_I) & = & \Phi_{n}^g(z;z_I) + \Res_{z \rightarrow \pm 1} \frac{\Delta_{z} G(z_0,z)}{4}\,\Delta_{z}\omega_n^g(z,z_I) \\
\label{Phind}\Phi_n^g(z_0;z_I) & = & \frac{1}{4{\rm i}\pi}\oint_{z \in \mathbb{U}} G(z_0,z)\,\dd_{z}\mathcal{V}_{n}^g(z;z_I) = \frac{1}{4{\rm i}\pi}\oint_{z \in \mathbb{U}} \omega_2^0(z_0,z)\,\mathcal{V}_n^g(z;z_I).
\eea
The expression \eqref{Phind} is valid when $z_0$ is outside $\mathbb{U}$, and can be analytically continued inside $\mathbb{U}$. We remind that $\Phi_n^{g}(z,z_I)$ is holomorphic in a neighborhood of $\mathbb{U}$. Then, we decompose $\mathcal{Q}_n^g$ defined in \eqref{Qde} as:
\bea
\label{recas}\mathcal{Q}_{n}^g(z;z_I) & = & \frac{1}{2}\,\mathcal{S}_{z}\omega_1^0(z)\,\mathcal{S}_{z}\omega_n^g(z,z_I) + \frac{1}{2}\,\Delta_{z}\omega_1^0(z)\,\Delta_{z}\omega_n^g(z,z_I) + \mathcal{E}_{n}^g(z;z_I) \\
\mathcal{E}_n^g(z;z_I) & = & \omega_{n + 1}^{g - 1}(z,\iota(z),z_I) + \sum_{\substack{J \subseteq I,\,\,0 \leq f \leq g \\ (J,f) \neq (\emptyset,0),(I,g)}} \omega_{|J| + 1}^{f}(z,z_{J})\,\omega_{n - |J|}^{g - f}(\iota(z),z_{I\setminus J}). 
\eea
The first term in \eqref{recas} has a double zero at $z = \pm 1$. So does $\mathcal{Q}_n^g(z;z_I)$ according to Theorem~\ref{thet2}. Therefore, if we plug the expression for $\Delta_{z}\mathcal{W}_n^g(z,z_I)$ in terms of $\mathcal{Q}_n^g(z,z_I)$, we find that the only term contributing to the residue in \eqref{repsa} is $\mathcal{E}_n^g(z,\iota(z),z_I)$. So, we have proved:
\begin{theorem}
\label{tpoth}
If $t,\mathbf{t}$ is tame, we have the recursion relation, for any $(n,g) \neq (1,0),(2,0)$:
\beq
\label{recto}\omega_n^g(z_0,z_I) = \Phi_{n}^g(z;z_I) + \Res_{z \rightarrow 1} K(z_0,z)\Big[\omega_{n + 1}^{g - 1}(z,\iota(z),z_I) + \sum_{\substack{J \subseteq I,\,\,0 \leq f \leq g \\ (J,h) \neq (\emptyset,0),(I,g)}} \omega_{|J| + 1}^{f}(z,z_J)\,\omega_{n - |J|}^{g - f}(\iota(z),z_{I\setminus J})\Big].
\eeq
\end{theorem}
This is a topological recursion, since the right-hand side involves only $\omega_{n'}^{g'}$ with $2g' - 2 + n' < 2g - 2 + n$. 
The form of the recursion is universal, it only depends on the model through the initial condition $\omega_1^0$ and $\omega_2^0$, and the monodromy operator $\iota$. Evaluating $\omega_{n'}^{g'}(z_1,z_J)$ at $z_1 = \iota(z)$ is done by Theorem~\ref{thet}, which led to the expression \eqref{thet3} for the monodromy.



\subsection{Examples of Euler characteristics $-1$}

\subsubsection{Torus with $1$ boundary}

For $(n,g) = (1,1)$, \eqref{recto} becomes:
\beq
\omega_1^1(z_0) = \Phi_{1}^1(z) + \Res_{z \rightarrow \pm 1} K(z_0,z)\,\omega_2^0(z,\iota(z)),
\eeq
and \eqref{Phind} gives:
\bea
\Phi_1^1(z_0) & = & \frac{1}{4{\rm i}\pi}\,\frac{2}{(k - 3)!} \sum_{k \geq 3} \int_{\mathbb{U}^{k}} 2\,T_k^0(x(\zeta_1),\ldots,x(\zeta_k))\,\omega_2^0(z_0,\zeta_1)\,\omega_2^0(\zeta_2,\zeta_3) \prod_{j = 4}^k \omega_1^0(\zeta_j) \nonumber \\
& & + \frac{1}{4{\rm i}\pi} \frac{1}{(k - 1)!} \sum_{k \geq 1} \int_{\mathbb{U}^k} 2\,T_k^1(x(\zeta_1),\ldots,x(\zeta_k))\,\omega_2^0(z_0,\zeta_1)\,\prod_{j = 2}^k \omega_1^0(\zeta_j).
\eea

\subsubsection{Sphere with $3$ boundaries}

For $(n,g) = (3,0)$, we compute from \eqref{recto}:
\bea
\omega_3^0(z_1,z_2,z_3) & = & \Phi_{3}^0(z_1;z_2,z_3) + \Res_{z \rightarrow \pm 1} K(z_1,z)\Big(\omega_{2}^0(z,z_2)\omega_2^0(\iota(z),z_3) + \omega_2^0(z,z_3)\omega_2^0(\iota(z),z_2)\Big) \nonumber \\
\label{intea}& = & \Res_{z \rightarrow \pm 1} \frac{\omega_2^0(z,z_1)\omega_2^0(z,z_2)\omega_2^0(z,z_3)}{2\dd x(z)\,\dd y(z)},
\eea
where we have defined the function $y$ which is the analytic continuation of $\Delta_{x}W_1^0(x)$ in the $z$-plane, and has simple zeroes at $z = \pm 1$. The integrand in \eqref{intea} has a simple pole at $z = \pm 1$ owing to $\dd x(z)$ in the denominator. Hence, the residue can be evaluated:
\beq
\frac{\omega_2^0(\underline{1},z_1)\omega_2^0(\underline{1},z_2)\omega_2^0(\underline{1},z_3)}{x'(1)y'(1)} + \frac{\omega_2^0(\underline{-1},z_1)\omega_2^0(\underline{-1},z_2)\omega_2^0(\underline{-1},z_3)}{x'(-1)y'(-1)},
\eeq
where $\underline{\alpha}$ means that we divide the $1$-form by $\dd z$ and evaluate the function obtained in this way at $z = \alpha$. Besides, from \eqref{Phind}, we have:
\beq
\Phi_3^0(z_1,z_2,z_3) = \frac{1}{4{\rm i}\pi}\sum_{k \geq 3} \frac{1}{(k - 3)!} \oint_{\mathbb{U}^k} T_k^0(x(\zeta_1),\ldots,x(\zeta_k))\,\omega_2^0(\zeta_1,z_1)\,\omega_2^0(\zeta_2,z_2)\,\omega_2^0(\zeta_3,z_3)\,\prod_{j = 4}^k \omega_1^0(\zeta_j).
\eeq
We observe that both $\omega_3^0(z_1,z_2,z_3)$ and $\Phi_3^0(z_1,z_2,z_3)$ are symmetric in their $3$ variables, although this is not obvious of the definition.

We leave to a future investigation the study of the symmetry properties of $\omega_n^g(z_1,\ldots,z_n)$ and $\Phi_n^g(z_1,\ldots,z_n)$.

\subsection{Generating series of closed stuffed maps}

The generating series of connected closed stuffed maps of genus $g$ is denoted $F^g$ (see \eqref{212}). It is characterized by its derivatives with respect to the parameters $\mathbf{t}$ of the model:
\beq
\label{resqu}\frac{\partial F^g}{\partial t_{m_1,\ldots,m_k}^h} = (-1)^k\,\Res_{x_1 \rightarrow \infty} \cdots \Res_{x_n \rightarrow \infty} \Big[\prod_{i = 1}^k x_i^{m_i}\dd x_i\Big]\Big(\!\!\!\!\!\!\!\!\!\!\!\sum_{\substack{K \vdash \ldbrack 1,k \rdbrack \\ f_1,\ldots,f_{[K]} \geq 0 \\ h + (\sum_i f_i) + k - [K] = g}} \!\!\!\!\!\!\!\!\!\!\!\! W_{|K_i|}^{f_i}(x_{K_i})\Big).
\eeq
The residue just picks up the coefficient of $x^{-(m_1 + 1)}\cdots x_k^{-(m_k + 1)}$ in the Laurent expansion at $\infty$ of the integrand. We leave to a future investigation the simultaneous integration of \eqref{resqu} to get a closed formula for $F^g$ in terms of $W_{n'}^{g'}$'s. For usual maps, this step was performed systematically in \cite{CE05}, but the problem here seems more complicated since the evaluation of $\omega_n^g(z_1,\ldots)$ at $z_1 = \iota(z)$ involves the operator $\mathcal{O}$ in \eqref{Oop} and thus depends explicitly on $\mathbf{t}$.

\subsection{Abstract loop equations with initial conditions}

In the terminology of \cite{BEO}, Theorem~\ref{thet} means that $\omega_{\bullet}^{\bullet}$ defined by \eqref{de21}-\eqref{de22}-\eqref{Phind} satisfy linear loop equations, which are here solvable thanks to Lemma~\ref{L36} because we work in the realm of formal series in $t,\mathbf{t}$. Theorem~\ref{thet2} then established that $\omega_{\bullet}^{\bullet}$ satisfies quadratic loop equation. The recursion formula \eqref{recto} is then shown in \cite[Proposition 2.7]{BEO} to be a consequence of those two properties, \S~\ref{S66} merely follows the proof of this result.

For usual maps (or for the usual $1$-hermitian matrix model), the relation between $W_n^g(x_1,\ldots,x_n)$ the generating series of maps (resp. the coefficients in a large $N$ expansion of the $n$-point correlation functions) and the $\omega_n^g$ satisfying the usual topological recursion of \cite{EOFg}, was:
\beq
\omega_n^g(z_1,\ldots,z_n) = W_n^g(x(z_1),\ldots,x(z_n)) + \delta_{n,2}\,\frac{\dd x(z_1)\,\dd x(z_2)}{\big(x(z_1) - x(z_2)\big)^2}\Big).
\eeq
It included a shift only for the unstable topologies $(n,g) = (1,0),(2,0)$. Here, for stuffed maps (or for the multi-trace hermitian matrix model), there is a shift between the residue formula and $\omega_n^g$ for any $(n,g)$, and this shift is given by $\Phi_n^g$ (see \eqref{Phind}), in terms of the potentials for topology $(n,g)$ discussed in \S~\ref{potS}. In some sense, we can see $\Phi_n^g$ as a way to include an "initial condition" for unstable topologies in the topological recursion.

\vspace{0.5cm}

\subsection*{Acknowledgments}

I thank B.~Eynard, E.~Guitter and N.~Orantin for asking questions which led to this project, the organizers of the Journ\'ees Cartes in June 2013 at the IPhT CEA Saclay where it was initiated, as well as S.~Garoufalidis, I.K.~Kostov and S.~Shadrin. This work is supported by the Max-Planck-Gesellschaft.

\appendix{}

\section{Two matrix model realization of stuffing}

Consider two $N \times N$ hermitian matrices with formal measure:
\beq
\label{w4}\dd\mu(M_1,M_2)\,\, \propto\,\, \dd M_1\,\dd M_2\,\mathrm{det}(1 - \alpha\,M_1\otimes M_2)^{-\gamma}\,\exp\big(-N\mathrm{Tr}\,V_1(M_1) - N\,\mathrm{Tr}\,V_2(M_2)\big).
\eeq
It induces on $M_1$ the distribution:
\bea
\dd\mu(M_1) &\,\,\propto\,\, & \dd M_1\,\exp\big(-N\,\mathrm{Tr}\,V_1(M_1)\big) \int_{\mathcal{H}_N} \dd M_2\,\exp\big(-N\,\mathrm{Tr}\,V_2(M_2)\big)\,\mathrm{det}(1 - \alpha M_1\otimes M_2)^{-\gamma} \\
& \propto & \dd M_1\,\exp\big(-N\,\mathrm{Tr}\,V_1(M_1)\big)\int_{\mathcal{H}_N} \dd M_2\,\exp\Big(- N\,\mathrm{Tr}\,V_2(M_2) + \gamma \sum_{\ell \geq 1} \frac{\alpha^\ell}{\ell}\,\mathrm{Tr}\,M_1^{\ell}\,\mathrm{Tr}\,M_2^{\ell}\Big) \nonumber \\
& \propto & \dd M_1\,\exp\big(-N\,\mathrm{Tr}\,V_1(M_1) + \sum_{k \geq 1} \frac{\gamma^k}{k!} \sum_{\ell_1,\ldots,\ell_k \geq 1} \frac{\check{T}_{\ell_1,\ldots,\ell_k}}{\ell_1\cdots\ell_k}\,\prod_{i = 1}^{k} \mathrm{Tr}\,M_1^{\ell_i}\Big),
\eea
where:
\beq
\check{T}_{\ell_1,\ldots,\ell_k} = \alpha^{\ell_1 + \cdots + \ell_k}\,\Big\langle \mathrm{Tr}\,M_2^{\ell_1}\cdots\mathrm{Tr}\,M_2^{\ell_k}\big\rangle_{M_2,c},
\eeq
and by definition:
\beq
\label{mues2}\langle f(M_2) \rangle_{M_2} = \frac{\int_{\mathcal{H}_N} \dd M_2\,\exp\big(-N\,\mathrm{Tr}\,V_2(M_2)\big)\,f(M_2)}{\int_{\mathcal{H}_N} \dd M_2\,\exp\big(- N\,\mathrm{Tr}\,V_2(M_2)\big)},
\eeq
and the subscript $c$ stands for "cumulant". In other words, the marginal distribution of $M_1$ in the model \eqref{w4} is of the form \eqref{e1}, where $T_{\ell_1,\ldots,\ell_k}$ are by definition the coefficients of the $k$-point correlators $\check{W}_k$ of the matrix $M_2$ for the measure defined in \eqref{mues2}:
\bea
\dd_{x_1}\cdots\dd_{x_k} \check{T}_k(x_1,\ldots,x_k) & = & \sum_{\ell_1,\ldots,\ell_k \geq 1} \check{T}_{\ell_1,\ldots,\ell_k}\,\prod_{i = 1}^k x^{\ell_i - 1} \dd x_i \\
& = & \check{W}_k\big(1/(\alpha x_1),\ldots,1/(\alpha x_k)\big)\dd\big(-1/(\alpha^2 x_1)\big)\cdots\dd\big(-1/(\alpha^2x_k)), \nonumber \\
\check{W}_k(\xi_1,\ldots,\xi_k) & = & \Big\langle \prod_{j = 1}^k \mathrm{Tr}\,\frac{1}{\xi_j - M_2}\Big\rangle_{M_2,c}.
\eea
The fatgraphs underlying the formal model \eqref{w4} are dual to usual maps with two types of faces (associated to $M_1$ or to $M_2$), and the particular coupling between $M_1$ and $M_2$ ensures that we can collect faces of the same type in clusters which are actually usual maps made of faces of type $M_1$ only. Therefore, a map appearing in the combinatorial description behind \eqref{w4} can be seen as a stuffed map (in the sense of \S~\ref{S21}) associated with $M_1$, whose elementary $2$-cells are themselves usual maps (of arbitrary topology) made of faces of type $M_2$. This justifies the name of "stuffing".

\newpage
\providecommand{\bysame}{\leavevmode\hbox to3em{\hrulefill}\thinspace}
\providecommand{\MR}{\relax\ifhmode\unskip\space\fi MR }
\providecommand{\MRhref}[2]{%
  \href{http://www.ams.org/mathscinet-getitem?mr=#1}{#2}
}
\providecommand{\href}[2]{#2}

\end{document}